\documentclass{article}

\usepackage{biorxiv} 

\usepackage[utf8]{inputenc} 
\usepackage[T1]{fontenc}    
\usepackage{hyperref}       
\usepackage{url}            
\usepackage{booktabs}       
\usepackage{amsfonts}       
\usepackage{nicefrac}       
\usepackage{microtype}      
\usepackage{lipsum}		
\usepackage{lineno}
\usepackage{algorithm}
\usepackage{arevmath}
\usepackage{amsmath}
\usepackage{amsthm}
\usepackage{graphicx}
\usepackage{tikz}
\usetikzlibrary{scopes,matrix,positioning}
\usepackage[noend]{algpseudocode}
\usepackage{caption}
\title{A Tutorial on the Mathematical Model of Single Cell Variational Inference}

\author{
  Songting Shi\\
  Department of Scientific and Engineering Computing\\ 
  School of Mathematical Sciences\\
  Peking University\\
  Beijing 300071, P. R. China \\
  \texttt{songtingstone@gmail.com} \\
  }

\begin{document}
\maketitle
\begin{abstract}
As the large amount of sequencing data accumulated in past decades and it is still accumulating, we need to handle the more and more sequencing data. As the fast development of the computing technologies, we now can handle a large amount of data by a reasonable of time using the neural network based model. This tutorial will introduce the the mathematical model of the single cell variational inference (scVI), which use the variational auto-encoder (building on the neural networks) to learn the distribution of the data to gain insights. It was written for beginners in the simple and intuitive way with many deduction details  to encourage more researchers into this field.  
\end{abstract}

As the computer technology evolves rapidly, we can tackle more and more complex problem by finding a suitable function taking millions of parameters to model the key part of the problems.  The deep neural network(\cite{DLarticle}, \cite{DLbook} is the top representer of such a function, and it has achieved success in the many fileds, such as natural language processing, image processing, game and so on, now it also gets into the computational biology, e.g. Alphafold(\cite{ Alphafold}).  This paper will introduce the single cell variational inference model scVI(\cite{scVI}), which use the variational auto-encoder equipped with the deep neural networks to tackle the data integration problem and downstream analysis on  the scRNA-seq data. 

The scVI model use the variational auto-encoder to extract information from the gene expression data.  The variational auto-encoder consists of the variational encoder and probability decoder, where the encoder will encode the expression data to a continuous hidden low dimensional space  in a compact form such that  cells from the  cell type will close to each other  even cells coming from different batches while different cell type separate; the decoder will decode the "code" in the common low dimensional space into the original space,
 it was designed to separate out the "dropout" effects in the sequencing to get the clean expressed data, which can be used to do imputation and  to find differential expressed genes. 
The "code"  in the low dimensional space of the cell can be used to do clustering, annotation, and visualization.
%

The basic idea of the variational auto-encoder is to learn the  distribution of the gene expression data by assume that the expression data was generated by two staged processes, the first stage is  to sample a code (may be view as the identifier of the cell) from the prior distribution on the low dimensional space, the second stage is to  sample a gene expression  from a conditional distribution based on the code of the cell. It design a variational decoder to decode the gene expression of a cell into its code, and also a probability decoder to decode the code into the gene expression of the cell. The parameters of the probability distribution  of decoder and encoder were output by the deep neural networks. By approximating the log-likelihood by the variational lower lower bound which were calculated from the parameters of  the encoder and decoder, we can maximized the variational lower bound to approach the maximum log-likelihood of the observed data, which yields the approximate best probability decoder and encoder. Having the decoder and encoder, we can do the downstream analysis on the data, e.g.  clustering, annotation, and visualization, imputation, differential expressed genes and so on. 

To make an intuitive understand of the scVI model, we will introduce the auto-encoder in Section 1 and variation auto encoder in Section 2. If you are familiar with the auto-ender and variational auto-encoder, please go directly into the Section 3 for the mathematical model of scVI. 

\section{Auto-Encoder}
To understand the scVI model, we should first understand how the variational auto-encoder works.  And to make the  understanding the variational auto-encoder easier, we first introduce the auto-encoder(\cite{bengio2009learning}) which is a similar but  simple model. Now, let we think a simple example to get the ideas of auto-encoder. Suppose that  the hidden code $z \sim \text{Normal}(0,1)$ and  the data $x$ is generated by $x =g(z)=  [z, 2z]$. We have 
\begin{equation}
x \sim \text{Normal}([0,0], [1,2;2,4])
\end{equation}
formally, where the variance matrix of $x$
 \begin{equation}
 \Sigma_x = \left [
 \begin{array}{cc}
 1&2 \\
 2&4
 \end{array}
\right ] 
\end{equation} 
is not invertible, and $x$ is a degenerated normal distribution. 
Suppose that we see a set of samples of $x$,
\begin{equation}
X:=\left [
 \begin{array}{cc}
   -1.5 &-3 \\
   -0.5& -1\\
  -0.2&-0.4\\
  -0.1&-0.2 \\
  0&0\\
 0.1&0.2 \\
 0.2&0.4\\
 0.5& 1\\
 1 &2 \\
 \end{array}
 \right ] 
\end{equation} 
from the above generation process. While now suppose that  we only see the set of examples, we do not know the underline generation mechanism. We want to learn an encode (contraction) function $f(\cdot)$  such that we can represent $x$ in a compact form by $z=f(x)$ and also a decode function $g(\cdot)$ such that we can recover $x$ form its code $z$, i.e, $x=g(z)$. How can we do this?

By a simple linear regression, we can easily get the relation $x_2 = 2 x_1$. 
This means that $x$ lines on a one-dimensional manifold, we can easily find the the contraction function $z=f(x) = x_1$ and recovery the $x$ from code $z$ by the generation function $x = g(z) = [z, 2z]$. For this simple example, a simple guess solves this problem. Can we find an algorithm from the above process to formulate a general method to solve this kind of problem but with more complicated data? Yes! Auto-encoder is one of such a method. 
It is a framework to learn the generation function $g(\cdot)$ and   encode (contraction) function $f(\cdot)$. 
%
%
Now we apply the auto-encoder method  to go through this simple example to gain the basic ideas.
We now suppose that $g$ and  $f$ comes from the linear transform function, and $x$ lies on the one-dimension manifold, we can parametrize function  $f(x, a_1, a_2) = a_1 x_1 + a_2 x_2$ and $g(z) = [b_1z, b_2z]$. Then the auto-encoder will output $\tilde x := g \cdot f (x) = [b_1z, b_2 z]= [b_1(a_1 x_1 + a_2 x_2),b_2(a_1 x_1 + a_2 x_2)]$. The objective function of auto-encoder is given by 
\begin{equation}
 \frac 1 N \sum_{n = 1}^N ||\tilde x_n - x_n||^2
\end{equation}
\begin{equation}
\label{eq:loss-in-detail}
L(X, a_1,a_2, b_1, b_2) := \frac 1 N\sum_{n = 1}^N  (x_{n,1} - b_1(a_1 x_{n,1}  + a_2 x_{n,2} ))^2 +  (x_{n,2} - b_2(a_1 x_{n,1}  + a_2 x_{n,2} ))^2 
\end{equation}
And then using the SGD methods or its variants to train the model on the training data to minimize the objective function (\ref{eq:loss-in-detail} ). 

Question: Can we learn out the optimal solution $a_1 =1, a_2 = 0, b_1 =1, b_2 = 2$? or some other reasonable solution? yes!

Since this objection is differentiable, we can set the first order of the objective function to get the stable condition. 

\begin{equation}
 \begin{array}{l}
\frac {\partial L} { \partial a_1} = \frac 2 N \sum_{n = 1}^N  -b_1x_{n,1}  (x_{n,1} - b_1(a_1 x_{n,1}  + a_2 x_{n,2} )) -b_2 x_{n,1} (x_{n,2} - b_2(a_1 x_{n,1}  + a_2 x_{n,2} )) = 0 \\
\frac {\partial L} { \partial a_2} = \frac 2 N \sum_{n = 1}^N  -b_1x_{n,2}  (x_{n,1} - b_1(a_1 x_{n,1}  + a_2 x_{n,2} )) -b_2 x_{n,2} (x_{n,2} - b_2(a_1 x_{n,1}  + a_2 x_{n,2} )) = 0 \\
\frac {\partial L} { \partial b_1} = \frac 2 N \sum_{n = 1}^N  -(a_1 x_{n,1}  + a_2 x_{n,2} )  (x_{n,1} - b_1(a_1 x_{n,1}  + a_2 x_{n,2} ))  = 0 \\
\frac {\partial L} { \partial b_2} = \frac 2 N \sum_{n = 1}^N  -(a_1 x_{n,1}  + a_2 x_{n,2} ) (x_{n,2} - b_2(a_1 x_{n,1}  + a_2 x_{n,2} )) = 0 
 \end{array}
\end{equation}
when we fix the $b_1, b_2$, we can get the following linear system about $a_1, a_2$.

\begin{equation}
\left [
 \begin{array}{cc}
  \sum_n (b_1^2 + b_2^2) x_{n,1}^2  &  \sum_n (b_1^2 + b_2^2) x_{n,1} x_{n,2} \\
  \sum_n (b_1^2 + b_2^2) x_{n,1} x_{n,2} & \sum_n (b_1^2 + b_2^2) x_{n,2}^2  \\
 \end{array}
 \right ] 
 \left [
 \begin{array}{c}
a_1 \\
a_2  \\
 \end{array}
 \right ]  = 
  \left [
 \begin{array}{c}
\sum_n  b_1 x_{n,1}^2 + b_2 x_{n,1} x_{n,2} \\
\sum_n  b_1x_{n,1} x_{n,2} + b_2  x_{n,1}^2 \\
 \end{array}
 \right ]
\end{equation}
Note that $x_{n,2} = 2x_{n,1}$, we can simplify it to the follow equation
\begin{equation}
\left [
 \begin{array}{cc}
 (b_1^2 + b_2^2)  &2(b_1^2 + b_2^2) \\
  2 (b_1^2 + b_2^2) &4 (b_1^2 + b_2^2)  \\
 \end{array}
 \right ] 
 \left [
 \begin{array}{c}
a_1 \\
a_2  \\
 \end{array}
 \right ]  = 
  \left [
 \begin{array}{c}
 b_1  + 2 b_2 \\
  2b_1 + 4 b_2   \\
 \end{array}
 \right ]
\end{equation}
Note that in the above equation, the second equation is a double times of the first equation, we simply get 
\begin{equation}
(b_1^2 + b_2^2) (a_1 + 2a_2) = b_1 + 2 b_2 \\
\end{equation}

When we fix $a_1, a_2$, we can get the following solution of $b_1, b_2$, 
\begin{equation}
 \begin{array}{c}
b_1 = \frac {\sum_n x_{n,1} (a_1x_{n,1} + a_2x_{n,2})} {\sum_n (a_1x_{n,1} + a_2 x_{n,2})^2} \\
b_2 = \frac {\sum_n x_{n,2} (a_1x_{n,1} + a_2x_{n,2})} {\sum_n (a_1x_{n,1} + a_2 x_{n,2})^2} \\
 \end{array}
\end{equation}
Note that $x_{n,2} = 2x_{n,1}$, we can simplify it to the follow equation
\begin{equation}\
 \begin{array}{c}
b_1 = \frac {\sum_n (a_1 + 2a_2)} {\sum_n (a_1+ 2a_2 )^2} \\
b_2 = \frac {\sum_n 2 (a_1 + 2a_2)}  {\sum_n (a_1+ 2a_2 )^2} = 2b_1 \\
 \end{array}
\end{equation}
Bring them  together, we get the following necessary condition of the stable point. 
\begin{equation}\
 \begin{array}{c}
 (b_1^2 + b_2^2) (a_1 + 2a_2) = b_1 + 2 b_2 \\
b_1 = \frac {(a_1 + 2a_2)} { (a_1+ a_2 )^2} \\
b_2 = \frac { 2 (a_1 + 2a_2)}  { (a_1+ a_2 )^2} = 2b_1 \\ 
 \end{array}
\end{equation}
It can be simplifed to the following equation

\begin{equation}\
 \begin{array}{c}
 b_1^2 (a_1 + 2a_2) = b_1  \\
b_1 = \frac { (a_1 + 2a_2)} {(a_1+ 2a_2 )^2} \\
b_2  = 2b_1 \\ 
 \end{array}
\end{equation}
When we restrict that $a_1 + 2a_2 \ne 0, b_1\ne 0$, we can get 
\begin{equation}
\label{eq:stable-condition}
 \begin{array}{c}
 b_1 (a_1 + 2a_2) =1  \\
b_2  = 2b_1 \\ 
 \end{array}
\end{equation}
Obviously, $a_1 =1, a_2 = 0, b_1 =1, b_2 = 2$ satisfy the stable condition (\ref{eq:stable-condition}). Also note that there are infinite solution of equation (\ref{eq:stable-condition}), e,g. , $a_1 =0, a_2 = 0.5, b_1 =1, b_2 = 2$ , $a_1 =0, a_2 = 0.25, b_1 =2, b_2 = 4$, and so on. And which solution arrived is depends on the which  algorithm been used. As in the variational auto-encoder, we restrict that $z$ approaches the standard normal distribution, which restrict that the $z$ has zero mean and unit variance, then we can get $a_1+ 2a_2 = 1, b_1=1, b_2=2$. Note that even in this case, there is a freedom in $a_1+ 2a_2 = 1$, but it do not influence the output $z = a_1 x_1 + a_2 x_2 = (a_1 + 2 a_2) x_1= x_1$.

When there is a freedom of the optimal parameters in the function, it usually will cause the optimization algorithm unstable since it can jump between the many optimums. If there are many parameters of function than which need to fit the true solution, it will cause overfitting of the training data which we learning the noise information in the training data in the function which will deviate  from  the true solution.  
A general principle is  to add a penalty on the objective function to avoid it, and the penalty can the   l2/l1 norm of  the parameters of the function. We now add the l2 norm penalty on the parameters $a_1, a_2, b_1, b2$ with multiplier $\lambda$, it will give the following loss function:
\begin{equation}
 \frac 1 N \sum_{n = 1}^N ||\tilde x_n - x_n||^2 + \lambda  (a_1^2 + a_2^2 + b_1^2 + b_2^2)
\end{equation}
\begin{equation}
\label{eq:loss-in-detail-penalty-ae}
 \begin{array}{ll}
L(X, a_1,a_2, b_1, b_2, \lambda) :=& \frac 1 N\sum_{n = 1}^N  \{(x_{n,1} - b_1(a_1 x_{n,1}  + a_2 x_{n,2} ))^2 + \\
& (x_{n,2} - b_2(a_1 x_{n,1}  + a_2 x_{n,2} ))^2 \} + \lambda  (a_1^2 + a_2^2 + b_1^2 + b_2^2)
 \end{array}
\end{equation}
We carry out the same analysis above. First, we use the first order condition to get the following condition which  the parameter must obey when it arrives at a  local minimum of the objective function. 

\begin{equation}
 \begin{array}{l}
\frac {\partial L} { \partial a_1} = \{ \frac 2 N \sum_{n = 1}^N  -b_1x_{n,1}  (x_{n,1} - b_1(a_1 x_{n,1}  + a_2 x_{n,2} )) -b_2 x_{n,1} (x_{n,2} - b_2(a_1 x_{n,1}  + a_2 x_{n,2} )) \} + 2\lambda a_1 = 0 \\
\frac {\partial L} { \partial a_2} = \{ \frac 2 N \sum_{n = 1}^N  -b_1x_{n,2}  (x_{n,1} - b_1(a_1 x_{n,1}  + a_2 x_{n,2} )) -b_2 x_{n,2} (x_{n,2} - b_2(a_1 x_{n,1}  + a_2 x_{n,2} )) \} + 2\lambda a_2 = 0 \\
\frac {\partial L} { \partial b_1} = \{ \frac 2 N \sum_{n = 1}^N  -(a_1 x_{n,1}  + a_2 x_{n,2} )  (x_{n,1} - b_1(a_1 x_{n,1}  + a_2 x_{n,2} )) \}  +2 \lambda b_1  = 0 \\
\frac {\partial L} { \partial b_2} =\{ \frac 2 N \sum_{n = 1}^N  -(a_1 x_{n,1}  + a_2 x_{n,2} ) (x_{n,2} - b_2(a_1 x_{n,1}  + a_2 x_{n,2} )) \} +2 \lambda b_2  = 0 
 \end{array}
\end{equation}
when we fix the $b_1, b_2$, we can get the following linear system about $a_1, a_2$.

\begin{equation}
\left [
 \begin{array}{cc}
 \frac 1 N  \sum_n (b_1^2 + b_2^2) x_{n,1}^2  + \lambda &  \frac 1 N \sum_n (b_1^2 + b_2^2) x_{n,1} x_{n,2} \\
   \frac 1 N \sum_n (b_1^2 + b_2^2) x_{n,1} x_{n,2} &  \frac 1 N \sum_n (b_1^2 + b_2^2) x_{n,2}^2  + \lambda  \\
 \end{array}
 \right ] 
 \left [
 \begin{array}{c}
a_1 \\
a_2  \\
 \end{array}
 \right ]  = 
  \left [
 \begin{array}{c}
 \frac 1 N\sum_n  b_1 x_{n,1}^2 + b_2 x_{n,1} x_{n,2} \\
 \frac 1 N \sum_n  b_1x_{n,1} x_{n,2} + b_2  x_{n,1}^2 \\
 \end{array}
 \right ]
\end{equation}
Note that $x_{n,2} = 2x_{n,1}$, and denoting $\gamma := \frac 1 N \sum_n x_{n,1}^2, \; \delta := b_1^2 + b_2^2$, we can simplify it to the follow equation
\begin{equation}
\left [
 \begin{array}{cc}
\gamma \delta + \lambda &2\gamma \delta \\
 \gamma \delta  &4 \gamma \delta   + \lambda  \\
 \end{array}
 \right ] 
 \left [
 \begin{array}{c}
a_1 \\
a_2  \\
 \end{array}
 \right ]  = 
  \left [
 \begin{array}{c}
( b_1  + 2 b_2)\gamma \\
2( b_1  + 2 b_2)\gamma    \\
 \end{array}
 \right ]
\end{equation}
We now assume  $ \delta := b_1^2 + b_2^2 >0,\; \lambda >0$ which is true usually, so that the coefficient matrix is invertible. 
We have the following solution
\begin{equation}
 \left [
 \begin{array}{c}
a_1 \\
a_2  \\
 \end{array}
 \right ]  = 
 \frac {(b_1 + 2 b_2) \lambda}{(5 \gamma \delta + \lambda)\lambda} \left [
 \begin{array}{c}
 \gamma \\
2 \gamma   \\
 \end{array}
 \right ]
  = 
 \frac {b_1 + 2 b_2}{5 \gamma \delta + \lambda} \left [
 \begin{array}{c}
 \gamma \\
2 \gamma   \\
 \end{array}
 \right ]
\end{equation}

Note that if $\gamma$ is the sample estimation of $\mathbb E_{x_1 \sim \mathcal N(0,1) } x_1^2 = 1$, and if $b_1 = 1, b_2 = 2$ and $\lambda \approx 0$, then we have 
\begin{equation}
 \left [
 \begin{array}{c}
a_1 \\
a_2  \\
 \end{array}
 \right ]  \approx
 \left [
 \begin{array}{c}
\frac 1 5  \\
\frac 2 5   \\
 \end{array}
 \right ] 
\end{equation},
which results in  $a_1 + 2 a_2  \approx 1$, this is what we needed.  Note that when we add the l2 norm, we focus a optimum point form the original on a line to a point, the reason is that the l2 norm add a local convexity on the loss landscape. 

When we fix $a_1, a_2$, we can get the following solution of $b_1, b_2$, 
\begin{equation}
 \begin{array}{c}
b_1 = \frac {\frac 1 N \sum_n x_{n,1} (a_1x_{n,1} + a_2x_{n,2})} {\frac 1 N  \sum_n (a_1x_{n,1} + a_2 x_{n,2})^2 + \lambda } \\
b_2 = \frac {\frac 1 N  \sum_n x_{n,2} (a_1x_{n,1} + a_2x_{n,2})} {\frac 1 N  \sum_n (a_1x_{n,1} + a_2 x_{n,2})^2 + \lambda } \\
 \end{array}
\end{equation}
Note that $x_{n,2} = 2x_{n,1}$, we can simplify it to the follow equation
\begin{equation}\
 \begin{array}{c}
b_1 = \frac {(a_1 + 2a_2)\gamma} { (a_1+ 2a_2 )^2\gamma + \lambda} \\
b_2 = \frac {2 (a_1 + 2a_2) \gamma}  { (a_1+ 2a_2 )^2\gamma + \lambda} =2b_1 \\
 \end{array}
\end{equation}
Note that $\gamma \approx 1$, and if  $\lambda \approx 0$ and $a_1 \approx 1/5, a_2  \approx 2/5$, then we have $b_1 \approx 1, \; b_2 \approx 2$. 

Bring them  together, we get the following necessary condition of the stable point. 
\begin{equation}\
 \begin{array}{c}
a_1 = \frac {b_1 + 2b_2} { 5\gamma (b_1^2 + b_2^2) + \lambda} \gamma \\
a_2 = 2a_1 \\
b_1 = \frac {(a_1 + 2a_2)\gamma} { (a_1+ 2a_2 )^2\gamma + \lambda} \\
b_2 = 2b_1 \\ 
 \end{array}
\end{equation}
Under the condition that $\lambda >0$, we can solve the above equation to get the solutions 
 \begin{equation}\
 \begin{array}{c}
a_1 = \frac{ \sqrt{ 5 \sqrt \gamma - \lambda} } {5 \gamma}\\
a_2 = 2a_1 \\
b_1 = \frac {\sqrt{ 5 \sqrt \gamma - \lambda}}{5 \sqrt \gamma} \\
b_2 = 2b_1 \\ 
 \end{array}
\end{equation}
or 
 \begin{equation}\
 \begin{array}{c}
a_1 = - \frac{ \sqrt{ 5 \sqrt \gamma - \lambda} } {5 \gamma}\\
a_2 = 2a_1 \\
b_1 = - \frac {\sqrt{ 5 \sqrt \gamma - \lambda}}{5 \sqrt \gamma} \\
b_2 = 2b_1 \\ 
 \end{array}
\end{equation}
Now we focus on the positive solution, when $\lambda = 0, \gamma = 1$, we have 
 \begin{equation}\
 \begin{array}{c}
a_1 = \frac{ \sqrt{ 5 }} {5 }\\
a_2 = \frac{ 2\sqrt{ 5 }} {5 } \\
b_1 =  \frac{ \sqrt{ 5 }}{5 }\\
b_2 =\frac{ 2\sqrt{ 5 }} {5 } \\
 \end{array}
\end{equation}
For $x_2 = 2x_1$,  we have $z = f(x) = a_1x_1 + a_2 x_2 = (a_1 + 2 a_2) x_1 = 5a_1x_1=\sqrt 5 x_1$, and $g(z)=[b_1z, b_2z] = [\frac{ \sqrt{ 5 }}{5 } z, \frac{2 \sqrt{ 5 }}{5 }z] = [x_1,2x_1] = [x_1, x_2]$. This verifies the correctness of the solution. But in this case, we have  $z= \sqrt 5 x_1 \sim \mathcal N(0,5)$. 
 In generally, if we only have $x_2 = 2x_1$, then  $g(f(x)) = [b_1 (a_1x_1 + a_2x_2), b_2(a_1x_1 + a_2 x_2)] = [5a_1 b_1 x_1, 5 a_1b_1x_2] = [\frac {5\sqrt \gamma - \lambda} {5\gamma \sqrt \gamma} x_1,\frac {5\sqrt \gamma - \lambda} {5\gamma \sqrt \gamma} x_2]$.  and $z = 5a_1 x_1 = \frac{ \sqrt{ 5 \sqrt \gamma - \lambda} } { \gamma} x_1 \sim  \mathcal N(0, \frac{  5 \sqrt \gamma - \lambda } { \gamma^2} )$. 
In this case, when $\gamma \approx 1$ and $\lambda \approx 0$, we will recovery $x$ correctly. Under the l2 norm penalty, we reduce the infinite solution of original encoder and decoder to two solutions, and this will make the algorithm more stable, if we choose that a small $\lambda$, the optimum of the objective with l2 penalty will approximate one of the optimums of the original solution.  

What will happen if we apply the l1 penalty  to the original objective, we left the exploration to you. 

Note if we want to constrain the distribution of   $z$ to  a standard normal distribution, this will meet a obstacle, since the distribution of $z = f(x)$ depends on the distribution of $x$, i.e. if we known the distribution of $x$ is $p(x)$, then we can get the distribution of $z$ is $p(f^{-1}(z)) |\frac{\partial f^{-1}(z)} {\partial z}|$ when the $f(\cdot)$ is invertible and the 
determinant
$ |\frac{\partial f^{-1}(z)} {\partial z}|$  not  equal zero almost surely. But we do not known the probability distribution  of $x$, even we know the $p(x)$, $p(f^{-1}(z)) |\frac{\partial f^{-1}(z)} {\partial z}|$ when the $f(\cdot)$  is hard to compute so that we can not use the KL divergence between the distribution of $z$ and the normal distribution  to get  a penalty.  This yields the need of the variational auto-encoder. Before we give the story of it, we first summarize the above simple formulation of auto-encoder to the general auto-encoder.

In the general form of auto-encoder, such as use in the image processing, it  consists of encoder function $z= f(x, \theta_f)$ and decoder function $x = g(z, \theta_g)$ , they are represented by the  neural network with parameters $\theta_f, \theta_g$, respectively. And general form of the neural networks can be represented in the form $ f(x, \theta_f) = \sigma_m(A_m( \sigma_{m-1}(A_{m-1}( \cdots      \sigma_2(A_2\sigma_1(A_1x + {\mathbf b}_1) + {\mathbf b}_2)) + {\mathbf b}_{m-1}))  + {\mathbf b}_m)$, where $\sigma_i(\cdot)$ is an element-wise non-linear activation function(e.g., sigmoid, ReLU), $A_{i} \in \mathbb R^{ d_i \times d_{i-1}}$ is the linear projection matrix and ${\mathbf b}_{i} \in \mathbb R^{ d_i \times 1}$  is the intercept term, it has $m-1$ hidden layers and final $m$-th layer is the output layer, and parmameters $\theta_f:= \{A_1, \ldots, A_m, \;  \mathbf b_1, \ldots, \mathbf b_m\}$. 
The loss function  is given by 
\begin{equation}
 \frac 1 N \sum_{n = 1}^N ||g(f(x,\theta_f), \theta_g)- x_n||^2 + \lambda (||\theta_f||_2^2 + ||\theta_g||_2^2) 
\end{equation}
And it is optimized by the SGD algorithm or its variants, and these methods  only need the computation of the gradient of loss function on a mini-batch of samples essentially. 

\section{Variational Auto-Encoder}
Now, we begin to tell the general story of the variational auto-encoder(\cite{VAE}, \cite{TVAE}) with general symbol. After that we begin to introduce the scVI model, which is a variational encoder  designed for the scRNA-seq data. 

\begin{figure}
  \centering
  \includegraphics[width=120mm]{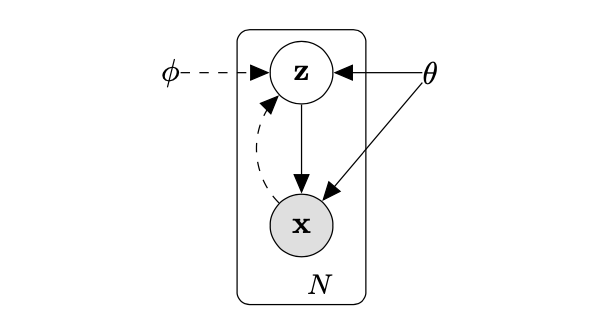}
   \caption{The type of directed graphical model under consideration. Solid lines denote the generative model $p_\theta(z) p_\theta(x|z)$, dashed lines denote the variational approximation $q_\phi(z|x)$ to the intractable posterior $p_\theta(z|x)$. The variational parameters $\phi$  are learned jointly with the generative model parameters $\theta$.}
\label{fig:generation-process} 
\end{figure}

We now use the same symbols in the \cite{VAE} to make it more easy to understand.  To tackle the uncomputable probability distribution of $z$, the variational auto-encoder assume that data point $x$ comes from the hidden continuous variable $z$. $z$ is generated from the probability distribution $p_{\theta^\star}(z)$, and then $x$ comes from the conditional distribution $p_{\theta^\star}(x|z)$. And this is represented in the Fig~\ref{fig:generation-process} (\cite{VAE})  with the solid arrow.

The probability distribution of $x$ is given by $p_{\theta} (x) = \int_z p_{\theta}(z)p_{\theta}(x|z)dz $. We hope that we can find a computable distribution $p_{\theta}(x|z), p_{\theta}(z|x), p_{\theta}(z)$ to concisely represented information from the data points $\{ x^{(i)},\; i=1,2, \cdots, N\}$, such that we can use these probability distribution to do the downstream analysis. As we known from the bayesian approach, we can use a probability class to represent the distribution $p_{\theta}(z), \; p_{\theta}(x|z)$, but the marginal distribution $p_{\theta} (x) $ is hard to obtain in general, so does the conditional distribution $p_{\theta}(z|x) = \frac{p_{\theta}(z)p_{\theta}(x|z)}{p_\theta(x)}$. The variational inference tackles this problem by using the computable distribution  $q_{\phi}(z|x)$ from the   distribution  class of  $p_\theta(z)$  to approximate the posterior distribution $p_{\theta}(z|x)$, which is represented by the dashed lines in Fig~\ref{fig:generation-process}. To achieve this goal, we need  find a computable algorithm to extract information form sample points into the parametrized distribution $p_{\theta}(z), \; p_{\theta}(x|z), q_{\phi}(z|x)$. This can finished by take the maximum likelihood method and do some approximation, i.e., use the variational lower bound. Now, we  begin to give the fundamental deduction of the variational lower bound.  
Firstly, in the classical maximum likelihood method, we seek the optimal $ \mathbf \theta^\star$ which maximize the log-likelihood 
$\log p_{ \mathbf \theta}(\mathbf x^{(1)}, \cdots, \mathbf x^{(N)}) = \sum_{i=1}^N \log p_{\mathbf \theta}(\mathbf x^{(i)})$.
The variational lower bound on the marginal likelihood of datapoint $i$ is defined by 
\begin{equation}\
 \begin{array}{c}
\mathcal L(\mathbf \theta,  \mathbf \phi, \mathbf x^{(i)} ):= \log p_{\mathbf \theta}(\mathbf x^{(i)}) - D_{KL}(q_{\mathbf \phi}(\mathbf z | \mathbf x^{(i)}) || p_{\mathbf \theta }(\mathbf z | \mathbf x^{(i)})) \\
 \end{array}
\end{equation}
The $D_{KL}(p(x) || q(x)):= \int_x p(x) \log \frac{p(x)}{ q(x)} dx$ is KL divergence between two distribution $p(x),\; q(x)$, which is nonnegative. The second RHS  term basically measure the divergence of approximate from the  true posterior. And since it is non-negative, we call it a lower bound.  We can rewrite the variational lower bound into the known quantities $p_{\theta}(z), \; p_{\theta}(x|z), q_{\phi}(z|x)$. 
\begin{equation}\
 \begin{array}{ll}
\mathcal L(\mathbf \theta,  \mathbf \phi, \mathbf x^{(i)} )&= \log p_{\mathbf \theta}(\mathbf x^{(i)}) - \mathbb E_{q_{\mathbf \phi}(\mathbf z | \mathbf x^{(i)}) } \log \frac {q_{\mathbf \phi}(\mathbf z | \mathbf x^{(i)}) }{ p_{\mathbf \theta }(\mathbf z | \mathbf x^{(i)})} \\
	&= \mathbb E_{q_{\mathbf \phi}(\mathbf z | \mathbf x^{(i)}) } [\log p_{\mathbf \theta}(\mathbf x^{(i)}) -  \log \frac {q_{\mathbf \phi}(\mathbf z | \mathbf x^{(i)}) }{ p_{\mathbf \theta }(\mathbf z | \mathbf x^{(i)})}] \\
	&= \mathbb E_{q_{\mathbf \phi}(\mathbf z | \mathbf x^{(i)}) } \log \frac   {  p_{\mathbf \theta}(\mathbf x^{(i)})  p_{\mathbf \theta }(\mathbf z | \mathbf x^{(i)})} {q_{\mathbf \phi}(\mathbf z | \mathbf x^{(i)}) }  \\
	&= \mathbb E_{q_{\mathbf \phi}(\mathbf z | \mathbf x^{(i)}) } \log \frac   {  p_{\mathbf \theta}(\mathbf z , \mathbf x^{(i)})} {q_{\mathbf \phi}(\mathbf z | \mathbf x^{(i)}) }  \\
	&= \mathbb E_{q_{\mathbf \phi}(\mathbf z | \mathbf x^{(i)}) } \log \frac   {  p_{\mathbf \theta}(\mathbf z)  p_{\mathbf \theta }(\mathbf x^{(i)} | \mathbf z) }  {q_{\mathbf \phi}(\mathbf z | \mathbf x^{(i)}) } \\
	&= - \mathbb E_{q_{\mathbf \phi}(\mathbf z | \mathbf x^{(i)}) } \log \frac  {q_{\mathbf \phi}(\mathbf z | \mathbf x^{(i)}) }   {  p_{\mathbf \theta}(\mathbf z)  }   + \mathbb E_{q_{\mathbf \phi}(\mathbf z | \mathbf x^{(i)}) }  \log p_{\mathbf \theta }(\mathbf x^{(i)} | \mathbf z)  \\
	&= -D_{KL}(q_{\mathbf \phi}(\mathbf z | \mathbf x^{(i)}) ||  p_{\mathbf \theta}(\mathbf z)  ) +  \mathbb E_{q_{\mathbf \phi}(\mathbf z | \mathbf x^{(i)}) }  \log p_{\mathbf \theta }(\mathbf x^{(i)} | \mathbf z)  \\
 \end{array}
\end{equation}
So we get the classical representation of the variational lower bound. 
\begin{equation}
\label{eq:varitional-lower-bound}
 \begin{array}{l}
\mathcal L(\mathbf \theta,  \mathbf \phi, \mathbf x^{(i)} )= -D_{KL}(q_{\mathbf \phi}(\mathbf z | \mathbf x^{(i)}) ||  p_{\mathbf \theta}(\mathbf z)  ) +  \mathbb E_{q_{\mathbf \phi}(\mathbf z | \mathbf x^{(i)}) }  \log p_{\mathbf \theta }(\mathbf x^{(i)} | \mathbf z)  \\
 \end{array}
\end{equation}

The first RHS term is the KL divergence between the approximate posterior $q_{\mathbf \phi}(\mathbf z | \mathbf x^{(i)})$ and the prior distribution $ p_{\mathbf \theta}(\mathbf z)$ of the hidden continuous variable $z$. When $q_{\mathbf \phi}(\mathbf z | \mathbf x^{(i)}) = p_{\mathbf \theta}(\mathbf z | \mathbf x^{(i)})$, we have a tight bound. 

\begin{equation}\
 \begin{array}{l}
\log p_{\mathbf \theta}(\mathbf x{(i)}) = \mathcal L(\mathbf \theta,  \mathbf \phi, \mathbf x^{(i)} )= -D_{KL}(p_{\mathbf \theta}(\mathbf z | \mathbf x^{(i)}) ||  p_{\mathbf \theta}(\mathbf z)  ) +  \mathbb E_{p_{\mathbf \theta}(\mathbf z | \mathbf x^{(i)}) }  \log p_{\mathbf \theta }(\mathbf x^{(i)} | \mathbf z)  \\
 \end{array}
\end{equation}
So if we fix the parameter $\mathbf \theta$, the maximum of the variational lower bound will equal the log-likelihood $p_{\mathbf \theta} (\mathbf x^{(i)})$, which is achieved by when $ p_{\mathbf \theta }(\mathbf x^{(i)} | \mathbf z)   = q_{\mathbf \phi}(\mathbf z | \mathbf x^{(i)})$. Now suppose that we always achieve such a state, i.e. the variational lower bound equals the marginal log-likelihood, by the maximum likelihood optimization, if we have large enough number of sample points, then the maximum of the log-likelihood will be achieved on the optimum $\theta^\star$. The above arguments roughly give us a belief that we can optimize the variational lower bound to find the optimum $\theta^\star$, and the $p_{\mathbf \theta^\star}(x)$ will catch up the underline data distribution. 

We next should select the proper distribution class with highly representative capacity  for the 
 distributions in the variational lower bound (\ref{eq:varitional-lower-bound}) to approximate the true distribution and make the optimization of the variational lower bound  easily and efficiently. 

Note that if $z$ is a continuous distribution in a $d$ dimensional space, e.g, normal distribution, $x$ is a random vector in the $d_2 \le d$ dimensional space, then we can find a function $f: \mathbb R^{d} \rightarrow \mathbb R^{d_2}$ such that $x=f(z)$ also surely(\cite{VAE}) with the proper complex function $f(\cdot)$. We can conjecture that if the random vector  $x \in \mathbb R^{n}, n\ge d_2$ lies on manifold with  essentially $d_2$ dimension, we can also find the function $f: \mathbb R^{d} \rightarrow \mathbb R^{n}$, such that $x = f(z)$. Now if $z \sim \mathcal N(\mathbf 0, \mathbf I)$, and $x$ is random vector represents the gene expression distribution. Since there are complicated regulatory network between genes, the function $f(z)=x$ should represent these complex regulatory networks. Now, the distribution of $z$ can be the simple normal distribution or log normal distribution, or other continuous distribution.  To make the KL divergence $D_{KL}(q_{\mathbf \phi}(\mathbf z | \mathbf x^{(i)}) ||  p_{\mathbf \theta}(\mathbf z)  )$ small, we let the approximate posterior $q_{\mathbf \phi}(\mathbf z | \mathbf x^{(i)})$ in the same distribution class of the distribution of  $z$. For the single cell RNA-seq data, the distribution class of $p_{\mathbf \theta }(\mathbf x^{(i)} | \mathbf z)$ choose the zero-inflated negative binomial distribution. 

We call the $q_{\mathbf \phi}(\mathbf z | \mathbf x^{(i)})$ as the encoder, it encoder the datapoint $ \mathbf x^{(i)}$ to its "code" $z$. And we refer $p_{\mathbf \theta }(\mathbf x^{(i)} | \mathbf z)$ as the decoder, it decode the "code" $z$ in the data point $ \mathbf x^{(i)}$. 

Here, we should point out that the complex regulatory networks between genes is modeled mainly by the mean of the negative binomial distribution. 
%
%

To get a sense of the final output by a independent Gassional variable with the mean and diagonal variance as a function of random variable $z$ will capture some dependence structure of $x$, we give a simple example. Now let $z\sim \mathcal N(0,1)$ is standard normal variable, and  $p(x|z)$  is the conditional density of $\mathcal N([z,2z], \text{diag}(1,1))$.  We can get the $p(x)$ in a close form. 
\begin{equation}
\begin{array}{ll}
p(x) 	&= \int p(z) p(x|z) dz\\
	&= \int \frac 1 {\sqrt{2\pi}} \exp(- \frac {z^2} 2) \frac 1 {2\pi} \exp (- \frac 1 2  ((x_1-z)^2 + (x_2 - 2z)^2)) dz \\
	&=\frac { 1}{ 2\pi \sqrt 6} \exp( -\frac 1 2 ( \frac 5 6 x_1^2 + \frac 1 3 x_2^2 - \frac 2 3 x_1 x_2) )
\end{array}
\end{equation}
So we get that 
\begin{equation}
x \sim \mathcal N \left ( 
\left [
\begin{array}{c}
0\\
0
\end{array}
\right ]
, 
\left [
\begin{array}{cc}
2 & 2\\
2 & 5
\end{array}
\right ]
\right )
\end{equation}
It shows that this simple example will capture the dependence of $x_1,x_2$ with $\text{Cov}(x_1, x_2) =2 = \text{Cov}(z, 2z) = \text{Cov}(\mu(z,\theta)_1,\mu(z,\theta)_2)$. So in general form $x \sim \mathcal N(\mathbf \mu(z,\theta), \text{diag}(\mathbf \sigma(z,\theta)))$ with the mean $\mathbf \mu$ and diagonal variance $\mathbf \sigma(z,\theta)$ output by nonlinear mapping such as neural networks, then the density $p(x)$ will capture complex dependence networks. If $x$ is the gene expressions, this will capture the complex  gene regulatory networks, and the complex  gene regulatory networks are captured by $\mathbf \mu(z,\theta)$. This may be one reason of the success of  the scVI model. 

The variational autoencoder(VAE) model the probability encoder $q_{\mathbf \phi}(\mathbf z | \mathbf x^{(i)})$ by modeling the parameters(i.e, the mean and diagonal covariance matrix) of the distribution with a nonlinear mapping (e.g. neural networks). $p_\theta(z)$ is the prior distribution usually the basic distribution without parameters $\theta$,  e.g. , standard Gauassion variables. And  $p_\theta(z)$  comes from the same probability distribution class  of  $q_{\mathbf \phi}(\mathbf z | \mathbf x^{(i)})$, this will lead a close form of the  KL divergence $D_{KL}(q_{\mathbf \phi}(\mathbf z | \mathbf x^{(i)}) ||  p_{\mathbf \theta}(\mathbf z) )$. The probability  distribution fo  probability decoder $ p_{\mathbf \theta }(\mathbf x^{(i)} | \mathbf z) $ should be accounts for the distribution of the real distribution of $x$, e.g. scVI choose the zero-inflated negative binomial distribution for the gene expression, while the image processing choose the Guassion distribution with diagonal variance. VAE use a nonlinear mapping ( neural networks) to model the parameters of the distribution of  $ p_{\mathbf \theta }(\mathbf x^{(i)} | \mathbf z)$. 

To train the neural networks on a large dataset, it use the stochastic optimization to train the model, which needs that a low variance estimate of the gradients of the objective function (variational lower bound). In most case, the parametric families of distribution of $p_{\theta}(z)$ will leads an analytical of expression $D_{KL}(q_{\mathbf \phi}(\mathbf z | \mathbf x^{(i)}) ||  p_{\mathbf \theta}(\mathbf z) )$  which is the differentiable with parameters $(\mathbf \theta, \mathbf \phi)$. While there is some problem with the reconstruction error term $  \mathbb E_{q_{\mathbf \phi}(\mathbf z | \mathbf x^{(i)}) }  \log p_{\mathbf \theta }(\mathbf x^{(i)} | \mathbf z) $ of the variational lower bound. If we use 
\begin{equation}
\begin{array}{ll}
\frac 1 L \sum_{l=1}^L  \log p_{\mathbf \theta }(\mathbf x^{(i)} | \mathbf z^{(i,l)}\\
\text{where  } \mathbf z^{(i,l)} \sim q_{\mathbf \phi}(\mathbf z | \mathbf x^{(i)})
\end{array}
\end{equation}
to estimate it, this will cause two problems. The first one is that variance of this estimation is very high, so it will fail the stochastic optimization. And the second one is that we can not differentiate it with parameters $\mathbf \phi$, since the backward gradient can not pass through a sample $ \mathbf z^{(i,l)}$  to the parameters of  the distribution  $q_{\mathbf \phi}(\mathbf z | \mathbf x^{(i)})$. To get around this problem, \cite{VAE} proposed  the reparametrization trick. The trick use the fact that we can express the random variable  $ \mathbf z \sim q_{\mathbf \phi}(\mathbf z | \mathbf x)$ by a deterministic function $z = f_{\mathbf \phi}(\mathbf \epsilon, \mathbf x)$   in many cases, where $\mathbf \epsilon$ is auxiliary random variable with a independent marginal distribution $p(\mathbf \epsilon)$. For example $z\sim \mathcal N(\mu,\sigma^2)$ can be expressed by $z = \mu + \sigma \epsilon, \; \epsilon \sim \mathcal N(0,1)$.  As we known that $q_{\mathbf \phi}(\mathbf z | \mathbf x^{(i)}) \prod_i dz_i = p(\mathbf \epsilon) \prod_i d\epsilon_i$, so we have $ \mathbb E_{q_{\mathbf \phi}(\mathbf z | \mathbf x^{(i)}) }  \log p_{\mathbf \theta }(\mathbf x^{(i)} | \mathbf z) =  \mathbb E_{\mathbf \epsilon \sim p(\mathbf \epsilon)} \log p_{\mathbf \theta }(\mathbf x^{(i)} |f_{\mathbf \phi}(\mathbf \epsilon, \mathbf x^{(i)}) )$, which can be estimated by 
\begin{equation}
\begin{array}{ll}
\mathbb E_{\mathbf \epsilon \sim p(\mathbf \epsilon)} \log p_{\mathbf \theta }(\mathbf x^{(i)} |f_{\mathbf \phi}(\mathbf \epsilon, \mathbf x^{(i)}) ) \approx \frac 1 L \sum_{l=1}^L \log  p_{\mathbf \theta }(\mathbf x^{(i)} |  f_{\mathbf \phi}(\mathbf \epsilon^{(i,l)}, \mathbf x^{(i)}) ) \\ 
\text{where  } \mathbf \epsilon^{(i,l)} \sim p(\mathbf \epsilon)
\end{array}
\end{equation}
Now this estimate is differentiable with parameters $\mathbf \phi$. The variance of the this estimate is lower since $p(\epsilon)$ is an independent distribution which is not evolved with $\mathbf x,\;\mathbf \phi $ and it much easier to draw samples from the stationary distribution $p(\epsilon)$ to cover the the probability area than draw the same number samples from $q_{\mathbf \phi}(\mathbf z | \mathbf x^{(i)})$. We can sample only one point ($L=1$) due to the low variance of this estimate in many cases. 

Summarizing the above efforts, we approximate the log-likelihood by the variational lower bound, and we reparameterize  the reconstruction term of variational lower bound to get an equivalent representation, 
\begin{equation}
\label{eq:varitional-lower-bound-2}
 \begin{array}{ll}
\mathcal L(\mathbf \theta,  \mathbf \phi, \mathbf x^{(i)} )&= -D_{KL}(q_{\mathbf \phi}(\mathbf z | \mathbf x^{(i)}) ||  p_{\mathbf \theta}(\mathbf z)  ) +  \mathbb E_{q_{\mathbf \phi}(\mathbf z | \mathbf x^{(i)}) }  \log p_{\mathbf \theta }(\mathbf x^{(i)} | \mathbf z)  \\
	&= -D_{KL}(q_{\mathbf \phi}(\mathbf z | \mathbf x^{(i)}) ||  p_{\mathbf \theta}(\mathbf z)  ) +   \mathbb E_{\mathbf \epsilon \sim p(\mathbf \epsilon)} \log p_{\mathbf \theta }(\mathbf x^{(i)} |f_{\mathbf \phi}(\mathbf \epsilon, \mathbf x^{(i)}) )  \\
	& \text{where }  \mathbf z =f_{\mathbf \phi}(\mathbf \epsilon, \mathbf x^{(i)})  \sim q_{\mathbf \phi}(\mathbf z | \mathbf x^{(i)}), \;  \mathbf \epsilon \sim p(\mathbf \epsilon)
 \end{array}
\end{equation}
The equivalent representation of the variational lower bound is approximated by 
\begin{equation}
\label{eq:varitional-lower-bound-3}
 \begin{array}{ll}
 \mathcal L(\mathbf \theta,  \mathbf \phi, \mathbf x^{(i)} )&\approx\\
\tilde{\mathcal L}(\mathbf \theta,  \mathbf \phi, \mathbf x^{(i)} )&:= -D_{KL}(q_{\mathbf \phi}(\mathbf z | \mathbf x^{(i)}) ||  p_{\mathbf \theta}(\mathbf z)  ) +  
   \frac 1 L \sum_{l=1}^L \log  p_{\mathbf \theta }(\mathbf x^{(i)} |  f_{\mathbf \phi}(\mathbf \epsilon^{(i,l)}, \mathbf x^{(i)}) ) \\ 
&\text{where  } \mathbf \epsilon^{(i,l)} \sim p(\mathbf \epsilon)
\end{array}
\end{equation}
The variational auto-encoder will train the parameters of the neural networks  with the approximate  objective function  on a mini-batch of samples $ \sum_{i \in \text{mini-batch} }\tilde{\mathcal L}(\mathbf \theta,  \mathbf \phi, \mathbf x^{(i)} )$ each time to maximize the approximated log-likelihood with stochastic optimization methods, e.g. SGD, Adam(\cite{Adam}), and so on. And it is hopefully that the final solution output by the algorithm will approach the true optimum point $(\mathbf \theta^\star, \mathbf \phi^\star)$.

Now we return to  the  simple example above  to check the power of the variational auto-encoder. Suppose that the $x \in \mathbb R^2$ comes from the following generation process. 
\begin{equation}
\begin{array}{ll}
z \sim \mathcal N(0,1)\\
x \sim \mathcal N \left (
\left [
\begin{array}{c}
z\\
2z
\end{array}
\right ], 
\left [
\begin{array}{cc}
0 &1 \\
1 & 0
\end{array}
\right ]
\right )
\end{array}
\end{equation}
We have the follow probability density function. 
\begin{equation}
\begin{array}{ll}
p_{\theta^\star}(z) &= \frac 1 {\sqrt{2\pi}} \exp( - \frac {z^2} 2)  \\
p_{\theta^\star}(x|z) &= \frac 1 {2 \pi} \exp(-\frac {(x_1 - z)^2 + (x_2 -2z)^2} 2) \\
p_{\theta^\star}(x) & = \frac { 1}{ 2\pi \sqrt 6} \exp( -\frac 1 2 ( \frac 5 6 x_1^2 + \frac 1 3 x_2^2 - \frac 2 3 x_1 x_2) ) \sim  \mathcal N \left ( 
\left [
\begin{array}{c}
0\\
0
\end{array}
\right ]
, 
\left [
\begin{array}{cc}
2 & 2\\
2 & 5
\end{array}
\right ]
\right )\\
p_{\theta^\star}(z|x) &= \frac{p_{\theta^\star}(x|z) p_{\theta^\star}(z) }{p_{\theta^\star}(x)}  \\
	&= \frac{1} {\sqrt{ \frac{2\pi} 6}} \exp(-\frac{ (z- \frac{x_1 + 2x_2} 6 )^2} {2*\frac 1 6} ) \sim  \mathcal N(\frac{x_1 + 2x_2} 6, \frac 1 6)\\
\end{array}
\end{equation}
We now choose that  $q_\phi(z|x) \sim \mathcal N(\mu(x,\phi), \sigma^2(x,\phi))$ and $p_\theta(x|z) \sim \mathcal N( \mu(z,\theta), \text{diag}(\sigma^2(z,\theta)))$, where $\mu(x,\phi) \in \mathbb R,\; \sigma(x,\phi) \in \mathbb R_{+}$ are the function of $x$ with parameters $\phi$, and  $\mu(z,\theta) \in \mathbb R^2,\; \sigma(z,\theta) \in \mathbb R^2_{+}$  are the mapping of variable $z$ with parameters $\theta$. 

Chosen the model in the Gaussion classes, we can calculate the variational lower bound with analytical expression.
\begin{equation}
\begin{array}{ll}
\mathcal L(\mathbf \theta,  \mathbf \phi, \mathbf x )&= -D_{KL}(q_{\mathbf \phi}(\mathbf z | \mathbf x ||  p_{\mathbf \theta}(\mathbf z)  ) +  \mathbb E_{q_{\mathbf \phi}(\mathbf z | \mathbf x) }  \log p_{\mathbf \theta }(\mathbf x | \mathbf z)  \\
	&= -D_{KL} ( \mathcal N(\mu(x,\phi), \sigma^2(x,\phi)) || \mathcal N(0,1) ) + \mathbb E_{z \sim \mathcal N(\mu(x,\phi), \sigma^2(x,\phi)) } \log p_\theta(x|z) \\
	&= - [ - \log \sigma(x,\phi) - \frac 1 2 + \frac { \sigma^2(x,\phi) + \mu^2(x,\phi)}  2] \\
	& \quad +\mathbb E_{z \sim \mathcal N(\mu(x,\phi), \sigma^2(x,\phi)) } [- \log (2\pi \sigma_1(z,\theta) \sigma_2(z,\theta) ) - \frac {(x_1 - \mu_1(z,\theta))^2} {2 \sigma^2_1(z,\theta)} - \frac {(x_2 - \mu_2(z,\theta))^2} {2 \sigma^2_2(z,\theta)} ] \\
	&= - [ - \log \sigma(x,\phi) - \frac 1 2 + \frac { \sigma^2(x,\phi) + \mu^2(x,\phi)}  2] \\
	& \quad +\mathbb E_{\epsilon \sim \mathcal N(0,1) } [- \log (2\pi \sigma_1(\mu(x,\phi) + \sigma(x,\phi) \epsilon, \theta) \sigma_2( \mu(x,\phi) + \sigma(x,\phi) \epsilon ,\theta) ) \\
	&\quad - \frac {(x_1 - \mu_1( \mu(x,\phi) + \sigma(x,\phi) \epsilon ,\theta))^2} {2 \sigma^2_1( \mu(x,\phi) + \sigma(x,\phi) \epsilon ,\theta)} - \frac {(x_2 - \mu_2( \mu(x,\phi) + \sigma(x,\phi) \epsilon ,\theta))^2} {2 \sigma^2_2( \mu(x,\phi) + \sigma(x,\phi) \epsilon ,\theta)} ]
\end{array}
\end{equation}
To simplify the complex expression above, we suppose that $\sigma^2(x,\phi) = 1/6, \; \sigma^2_1(z, \theta) = 1,\; \sigma^2_2(z, \theta) = 1$, i.e we take the variance parameter the same as the underline true parameters. And we we get 
\begin{equation}
\begin{array}{ll}
\mathcal L(\mathbf \theta,  \mathbf \phi, \mathbf x )&=  - [ - \log \frac 1 {\sqrt{6}} - \frac 1 2 + \frac {  \frac 1 6 + \mu^2(x,\phi)}  2] \\
	& \quad +\mathbb E_{z \sim \mathcal N(\mu(x,\phi), \frac 1 6 ) } [- \log (2\pi  ) - \frac {(x_1 - \mu_1(z,\theta))^2} {2 } - \frac {(x_2 - \mu_2(z,\theta))^2} {2}] \\
	&= - \log (2\sqrt{6} \pi) + \frac 5 {12 } - \frac{\mu^2(x, \phi)} 2 - \mathbb E_{z \sim \mathcal N(\mu(x,\phi), \frac 1 6 ) } \frac {(x_1 - \mu_1(z,\theta))^2 + (x_2 - \mu_2(z,\theta))^2} 2
\end{array}
\end{equation}
We use the $\mu(x,\phi):= a_1x_1 + a_2x_2$, $\mu(z,\theta) := [b_1 z, b_2z]$  to parametrize the mean function, where $\phi = (a_1, a_2)$ and $\theta = (b_1, b_2)$. The above equation can be simplied to 
\begin{equation}
\begin{array}{ll}
\mathcal L(\mathbf \theta,  \mathbf \phi, \mathbf x )&=  - \log (2\sqrt{6} \pi) + \frac 5 {12 } - \frac {( a_1 x_1 + a_2 x_2)^2 } 2 - \frac {b_1^2 + b_2^2 } {12} - \frac{ (b_1(a_1x_1 + a_2 x_2) -x_1)^2 + (b_2(a_1x_1 + a_2 x_2) -x_2)^2} 2 
\end{array}
\end{equation}
And we can find the optimal solution of $\phi =(a_1, a_2 ), \; \theta=(b_1, b_2)$ with the following loss if we have samples $\{x_n, n=1, \ldots, N \} \sim \mathcal N \left ( 
\left [
\begin{array}{c}
0\\
0
\end{array}
\right ]
, 
\left [
\begin{array}{cc}
2 & 2\\
2 & 5
\end{array}
\right ]
\right )$.

\begin{equation}
\label{eq:loss-in-detail-penalty-vae-1}
\begin{array}{ll}
\underset {\phi,\theta}  {\min  } \;  L(\mathbf \theta,  \mathbf \phi)  =  \frac 1 N  \sum_{n=1}^N [ \frac {( a_1 x_{n1} + a_2 x_{n2})^2 } 2 + \frac {b_1^2 + b_2^2 } {12} + \frac{ (b_1(a_1x_{n1} + a_2 x_{n2}) -x_{n1})^2 + (b_2(a_1x_{n1}  + a_2 x_{n2}) -x_{n2})^2} 2 ]
\end{array}
\end{equation}

Comparing with the auto-encoder (AE) loss (\ref{eq:loss-in-detail-penalty-ae}),  the above variational auto-encoder(VAE) loss (\ref{eq:loss-in-detail-penalty-vae-1})  is similar to the auto-encoder loss. The VAE loss has one more term $z_n^2 = (a_1x_{n1} + a_2 x_{n2})^2$ than AE loss, this term comes from we want to make  that the probability encoder $q_\phi(z|x)$ close to the standard normal distribution, which is the term we want to do in the AE loss. Note that this inspire us we can add a penalty term $\beta \frac 1 N  \sum_n ||z_n||^2$ into the loss  (\ref{eq:loss-in-detail-penalty-ae}), which will bias to  a solution such that the $p_\theta(z)$  close to some normal distribution $\mathcal N(0, F(\beta, \lambda))$ where $F(\beta, \lambda)$ is a function of $\beta, \lambda$. We may get some satisfactory with this improvement.  However, we can not prespecify the coefficient $\beta, \lambda$ such that  $F(\beta, \lambda) =1$ while this can be accomplished in the variational auto-encoder, since it was deduced from the likelihood function. 
In this simplified case in which we assume that the the  $q_\phi(z|x)$ and $p_\theta(x|z)$ has a constant diagonal variance matrix. It gives us the intuition that the 
the mean $\mu(x,\phi)$  of probability  encoder $q_\phi(z|x)$ in  the VAE has a similar effect as the encoder function $z=f(x,\phi)$ in the AE, and the the mean $\mu(z, \theta)$  of probability  encoder $p_\theta(x|z)$ in  the VAE has a similar effect as the encoder function $z=g(x,\theta)$ in the AE. I make the conjecture that this fact is true when we have the probability encoder and decoder coming the Guassion class and has a diagonal variance matrix. The reason behind this phenomenon is that when we have a Gussion vector has a diagonal covariance, e.g.  $p_\theta(x|z) \sim \mathcal N( \mu(z,\theta), \text{diag}(\sigma^2(z,\theta)))$, the complex correlation between variables $x_g, \; g= 1, \ldots, G$ is mainly captured by the $\mu(z,\theta)$ since $x_g, \; g= 1, \ldots, G$ is independent when give the $z$, and if $z\sim \mathcal N(0, I)$, the $\mu(z,\theta)$ is a random vector and will capture the complex dependence between  variables $x_g, \; g= 1, \ldots, G$ via the nonlinear mapping function $\mu(z,\theta)$. 


Now we begin to solve the above optimization problem with first order condition. 
\begin{equation}
 \begin{array}{ll}
\frac {\partial L} { \partial a_1} &=\frac 1 N  \sum_{n=1}^N [ x_{n1} ( a_1 x_{n1} + a_2 x_{n2}) + b_1 x_{n1}(b_1(a_1x_{n1} + a_2 x_{n2}) -x_{n1}) \\
	&\quad + b_2 x_{n1} (b_2(a_1x_{n1}  + a_2 x_{n2}) -x_{n2}) ]
 = 0 \\
\frac {\partial L} { \partial a_2} &=\frac 1 N  \sum_{n=1}^N [ x_{n2} ( a_1 x_{n1} + a_2 x_{n2}) + b_1 x_{n2}(b_1(a_1x_{n1} 
	+ a_2 x_{n2}) -x_{n1})\\
	&\quad  + b_2 x_{n2} (b_2(a_1x_{n1}  + a_2 x_{n2}) -x_{n2}) ]
 = 0 \\
\frac {\partial L} { \partial b_1} &= \frac 1 N  \sum_{n=1}^N [ \frac {b_1 } {6} +  (a_1x_{n1} + a_2 x_{n2}) (b_1(a_1x_{n1} + a_2 x_{n2}) -x_{n1})  ]=0\\
\frac {\partial L} { \partial b_2} &=\frac 1 N  \sum_{n=1}^N [ \frac {b_2 } {6} +  (a_1x_{n1} + a_2 x_{n2}) (b_2(a_1x_{n1} + a_2 x_{n2}) -x_{n2})  ]= 0 
 \end{array}
\end{equation}
Now, we use the approximation $\frac 1 N \sum_{n=1}^N x_{n1}^2 \approx \mathbb E x_1^2 =2$,  $\frac 1 N \sum_{n=1}^N x_{n2}^2 \approx \mathbb E x_1^2 =5$,$\frac 1 N \sum_{n=1}^N x_{n1}  x_{n2} \approx \mathbb E x_1x_2 =2$ into the above equation, we get 
\begin{equation}
 \begin{array}{l}
( 2a_1 + 2a_2) + b_1(b_1(2a_1 + 2 a_2 ) - 2) + b_2  (b_2(2a_1 + 2a_2 ) -2 )
  \approx 0 \\
( 2a_1  + 5a_2 ) + b_1 (b_1(2a_1 + 5a_2 ) -2) + b_2  (b_2(2a_1  + 5a_2 ) -5) 
  \approx 0 \\
 \frac {b_1 } {6} +  b_1 (2a_1^2 + 5a_2^2 +4a_1a_2)  - 2(a_1 + a_2)  \approx 0\\
 \frac {b_2 } {6} +  b_2 (2a_1^2 + 5a_2^2 +4a_1a_2)  -(2a_1 + 5a_2) \approx 0 
 \end{array}
\end{equation}
I can not solve the above equation with explicit solution since it will evolve a five order equation about $a_1, a_2$ when eliminating $b_1,b_2$. But we can check that the optimal solution $\phi^\star = (a^\star_1, a^\star_2) = (1/6,, 1/3)$, $\theta^\star = (b^\star_1, b^\star_2) = (1, 2)$ satisfies the above stable condition exactly. This shows that the variational auto-encoder has the power to find the true solution. Note that when $x \sim  \mathcal N \left ( 
\left [
\begin{array}{c}
0\\
0
\end{array}
\right ]
, 
\left [
\begin{array}{cc}
2 & 2\\
2 & 5
\end{array}
\right ]
\right )$, 
the $\mu(x,\phi^\star) = 1/6 x_1 + 1/3 x_2 \sim \mathcal N(0, 5/6)$ is close to the true hidden variable distribution of $z \sim  \mathcal N(0, 1)$ but not the same, this is because that $z = \mu(x,\phi^\star)  + \epsilon,\; \epsilon \sim \mathcal N(0, 1/6)$. The $\mu(z,\theta^\star) = \left [
\begin{array}{c}
z\\
2z
\end{array}
\right ]
  \sim \mathcal N \left ( 
\left [
\begin{array}{c}
0\\
0
\end{array}
\right ]
, 
\left [
\begin{array}{cc}
1 & 2\\
2 & 4
\end{array}
\right ]
\right )$ is also close to the true distribution $ \mathcal N \left ( 
\left [
\begin{array}{c}
0\\
0
\end{array}
\right ]
, 
\left [
\begin{array}{cc}
2 & 2\\
2 & 5
\end{array}
\right ]
\right )$. 
If we set the  $ \mathcal N \left ( 
\left [
\begin{array}{c}
0\\
0
\end{array}
\right ]
, 
\left [
\begin{array}{cc}
2 & 2\\
2 & 5
\end{array}
\right ]
\right )$ as the true distribution of $x$, then we can re-tell the story. The $\tilde x = x + \epsilon_x$ is the measurements of $x \sim \mathcal N \left ( 
\left [
\begin{array}{c}
0\\
0
\end{array}
\right ]
, 
\left [
\begin{array}{cc}
1 & 2\\
2 & 4
\end{array}
\right ]
\right )$  with the measurement error $\epsilon_x \sim  \mathcal N \left ( 
\left [
\begin{array}{c}
0\\
0
\end{array}
\right ]
, 
\left [
\begin{array}{cc}
1 & 0\\
0 & 1
\end{array}
\right ]
\right )$, then we can use the variational auto-encoder method to approximate the distribution of $x$, which is given by distribution of the mean function $u(z,\theta^\star)$ of posterior $p_{\theta^\star}(x|z)$, i.e. $p(x) =  \int_{x= u(z,\theta^\star)} p(z)dz$. And this interpretation matches the  the data generation process in practice, such as the scRNA-seq data, the gene expression has the complex regulatory relations to control the protein production, while the measurement error may be independent for each gene, i.e. the expression data $\tilde x \in \mathbb R^G$  can be written  as $\tilde x = x + \epsilon_x$, where $x$ is true expression data, and $\epsilon_x$ is the measurement error, and genes $x_g, \; g=1, 2,  \ldots, G$ have complex regulatory relations, but the  $\epsilon_x(g),  \; g=1, 2,  \ldots, G$ are independent with each other and also are independent of $x$. So we can use the variational auto-encoder to model this process to assume that the distribution of measurements $\tilde x$ is comes from a parametric distribution $p_\theta(x|z)$ which can be characterized  by its mean and diagonal variance 
 when given the hidden continuous variable $z$ where the mean $\mu(z,\theta)$ of $p_\theta(x|z)$ give the distribution of $x$, i.e. $x= \mu(z,\theta)$, and the diagonal variance characterized the independent random error $\epsilon_x$. This phenomenon can go through when the $p_\theta(x|z)$ in the Gaussion class, i.e, $\tilde x \sim p_\theta(x|z) \sim \mathcal N( \mu(z,\theta), \text{diag}(\sigma^2(\theta))$, $\tilde x  = \mu(z,\theta) + \epsilon$ where $x= \mu(z,\theta)$ and  $\epsilon \sim  \mathcal N( 0, \text{diag}(\sigma^2(\theta)))$ is the independent random errors. 
When the distribution of $x$ is given by the negative binomial distribution $x \sim NB(dispersion= d(z,\theta), mu= \mu(z,\theta))$ in which each gene is independent of others when give $z$, the above intuition can roughly go through, we can use the $\mu(z,\theta)$ to capture the complex gene regulatory relation. But the variance which given by $\mu(z,\theta) + \frac{\mu^2(z,\theta)} {d(z,\theta)}$ also correlated for different genes, and there does not have an additive independent random noise which give the independent error for genes  in this kind of representation, it may be more resonable to find a discrete count distribution to model independent errors ( independent of the mean, also the errors are independent for different genes) in the measurement. 

Now, we generalize the above example a little. We replace the variance of the error of $1$ to $\gamma$ such that we can observe how the error do influence on the VAE method.

Suppose that the $x \in \mathbb R^2$ comes from the following generation process. 
\begin{equation}
\begin{array}{ll}
z \sim \mathcal N(0,1)\\
x = \left [
\begin{array}{c}
z\\
2z
\end{array}
\right ] \\
\epsilon \sim \mathcal N \left (
\left [
\begin{array}{c}
0\\
0
\end{array}
\right ], 
\left [
\begin{array}{cc}
0 &\gamma \\
\gamma & 0
\end{array}
\right ]
\right )\\
\tilde x = x + \epsilon
\end{array}
\end{equation}
where $z$ is he hidden continuous variable, $x$ is the random vector which we are interested in, $\epsilon$ is the measurement error which is independent of $z$. Note here the samples are comes from $\tilde x$ so that we should use the variational model on $\tilde x$, but we hope to denoise to get the distribution of $x$.

We have the follow probability density function. 
\begin{equation}
\begin{array}{ll}
p_{\theta^\star}(z) &= \frac 1 {\sqrt{2\pi}} \exp( - \frac {z^2} 2)  \\
p_{\theta^\star}(x|z) &=\delta(x_1-z) \delta(x_2-2z) \\
p_{\theta^\star}(\tilde x|z) &= \frac 1 {2 \pi \gamma} \exp(-\frac {(\tilde x_1 - z)^2 + (\tilde x_2 -2z)^2} {2 \gamma}) \\
p_{\theta^\star}(x) & = \frac { 1}{\sqrt{ 2\pi } } \exp( -\frac  {x_1^2} 2 ) \delta(x_2 - 2x_1)\\
p_{\theta^\star}(\tilde x) & = \frac { 1}{ 2\pi \sqrt {( 5 + \gamma ) \gamma }} \exp( -\frac  { (4 + \gamma) \tilde x_1^2 +  (1 + \gamma)  \tilde x_2^2 - 4 \tilde x_1 \tilde x_2} {2 \gamma (5 + \gamma)} ) \sim  \mathcal N \left ( 
\left [
\begin{array}{c}
0\\
0
\end{array}
\right ]
, 
\left [
\begin{array}{cc}
1 + \gamma & 2\\
2 &  4 + \gamma
\end{array}
\right ]
\right )\\
p_{\theta^\star}(z|x) &=\delta(z-x_1)\delta(z-x_2/2)\\
p_{\theta^\star}(z|\tilde x) &= \frac{p_{\theta^\star}(\tilde x|z) p_{\theta^\star}(z) }{p_{\theta^\star}(\tilde x)}  \\
	&= \frac{1} {\sqrt{ 2\pi  \frac \gamma { 5 + \gamma} }} \exp(-\frac{ (z- \frac{\tilde x_1 + 2 \tilde x_2} { 5 + \gamma } )^2} {2*\frac \gamma { 5+\gamma } } ) \sim  \mathcal N(\frac{\tilde  x_1 + 2\tilde  x_2} {5+\gamma}, \frac \gamma  {5 +\gamma} )\\
\end{array}
\end{equation}
where $\delta(\cdot)$ is the Dirac delta function, with the property $\delta(x):= \left \{
 \begin{array}{cl}
\infty & \text{if } x=0\\
0	& \text{otherwise}
\end{array}
\right.
$ and  $\int_{-x}^x \delta(t)dt =1, \forall x>0$.
We now choose that  $q_\phi(z|\tilde x) \sim \mathcal N(\mu(\tilde x,\phi), \sigma^2(\tilde x,\phi))$ and $p_\theta(\tilde x|z) \sim \mathcal N( \mu(z,\theta), \text{diag}(\sigma^2(z,\theta)))$, where $\mu(\tilde x,\phi) \in \mathbb R,\; \sigma(\tilde x,\phi) \in \mathbb R_{+}$ are the function of $\tilde x$ with parameters $\phi$, and  $\mu(z,\theta) \in \mathbb R^2,\; \sigma(z,\theta) \in \mathbb R^2_{+}$  are the mapping of variable $z$ with parameters $\theta$. 

Chosen the model in the Gaussion classes, we can calculate the variational lower bound with analytical expression.
\begin{equation}
\begin{array}{ll}
\mathcal L(\mathbf \theta,  \mathbf \phi, \mathbf{ \tilde x} )&= -D_{KL}(q_{\mathbf \phi}(\mathbf z | \mathbf{ \tilde x} ||  p_{\mathbf \theta}(\mathbf z)  ) +  \mathbb E_{q_{\mathbf \phi}(\mathbf z | \mathbf{ \tilde x}) }  \log p_{\mathbf \theta }(\mathbf{ \tilde x} | \mathbf z)  \\
	&= -D_{KL} ( \mathcal N(\mu(\tilde x,\phi), \sigma^2(\tilde x,\phi)) || \mathcal N(0,1) ) + \mathbb E_{z \sim \mathcal N(\mu(\tilde x,\phi), \sigma^2(\tilde x,\phi)) } \log p_\theta(\tilde x|z) \\
	&= - [ - \log \sigma(\tilde x,\phi) - \frac 1 2 + \frac { \sigma^2(\tilde x,\phi) + \mu^2(\tilde x,\phi)}  2] \\
	& \quad +\mathbb E_{z \sim \mathcal N(\mu(\tilde x,\phi), \sigma^2(\tilde x,\phi)) } [- \log (2\pi \sigma_1(z,\theta) \sigma_2(z,\theta) ) - \frac {(\tilde x_1 - \mu_1(z,\theta))^2} {2 \sigma^2_1(z,\theta)} - \frac {(\tilde x_2 - \mu_2(z,\theta))^2} {2 \sigma^2_2(z,\theta)} ] \\
	&= - [ - \log \sigma(\tilde x,\phi) - \frac 1 2 + \frac { \sigma^2(\tilde x,\phi) + \mu^2(\tilde x,\phi)}  2] \\
	& \quad +\mathbb E_{\epsilon_z \sim \mathcal N(0,1) } [- \log (2\pi \sigma_1(\mu(\tilde x,\phi) + \sigma(\tilde x,\phi) \epsilon, \theta) \sigma_2( \mu(\tilde x,\phi) + \sigma(\tilde x,\phi) \epsilon ,\theta) ) \\
	&\quad - \frac {(\tilde x_1 - \mu_1( \mu(\tilde x,\phi) + \sigma(\tilde x,\phi) \epsilon ,\theta))^2} {2 \sigma^2_1( \mu(\tilde x,\phi) + \sigma(\tilde x,\phi) \epsilon ,\theta)} - \frac {(\tilde x_2 - \mu_2( \mu(\tilde x,\phi) + \sigma(\tilde x,\phi) \epsilon ,\theta))^2} {2 \sigma^2_2( \mu(\tilde x,\phi) + \sigma(\tilde x,\phi) \epsilon ,\theta)} ]
\end{array}
\end{equation}
To simplify the complex expression above, we suppose that $\sigma^2(\tilde x,\phi) = \frac \gamma { 5 + \gamma }, \; \sigma^2_1(z, \theta) = \gamma,\; \sigma^2_2(z, \theta) = \gamma$, i.e we take the variance parameter the same as the underline true parameters. And we get 
\begin{equation}
\begin{array}{ll}
\mathcal L(\mathbf \theta,  \mathbf \phi, \mathbf {\tilde x} )&=  - [ - \frac 1 2 \log \frac { \gamma } { 5 + \gamma } -  \frac 1 2 + \frac {   \frac { \gamma } { 5 + \gamma } + \mu^2(\tilde x,\phi)}  2] \\
	& \quad +\mathbb E_{z \sim \mathcal N(\mu(\tilde x,\phi),   \frac { \gamma } { 5 + \gamma }  ) } [- \log (2\pi  \gamma ) - \frac {(\tilde x_1 - \mu_1(z,\theta))^2} {2 \gamma } - \frac {(\tilde x_2 - \mu_2(z,\theta))^2} {2 \gamma}] \\
	&= - \log (2 \pi \sqrt{ \gamma ( 5 + \gamma) })  + \frac 5 { 2(5 + \gamma)} -  \frac{\mu^2(\tilde x, \phi)} 2 - \mathbb E_{z \sim \mathcal N(\mu(\tilde x,\phi),  \frac { \gamma } { 5 + \gamma }  ) } \frac {(\tilde x_1 - \mu_1(z,\theta))^2 + (\tilde x_2 - \mu_2(z,\theta))^2} { 2 \gamma }
\end{array}
\end{equation}
We use the $\mu(\tilde x,\phi):= a_1\tilde x_1 + a_2\tilde x_2$, $\mu(z,\theta) := [b_1 z, b_2z]$  to parametrize the mean function, where $\phi = (a_1, a_2)$ and $\theta = (b_1, b_2)$. The above equation can be simplied to 
\begin{equation}
\begin{array}{ll}
\mathcal L(\mathbf \theta,  \mathbf \phi, \mathbf {\tilde x} )&= - \log (2 \pi \sqrt{ \gamma ( 5 + \gamma) })  + \frac 5 { 2(5 + \gamma)}  - \frac {( a_1 \tilde x_1 + a_2 \tilde x_2)^2 } 2 - \frac {b_1^2 + b_2^2 } {2 ( 5 + \gamma) } \\
	& \quad - \frac{ (b_1(a_1\tilde x_1 + a_2 \tilde x_2) -\tilde x_1)^2 + (b_2(a_1\tilde x_1 + a_2 \tilde x_2) -\tilde x_2)^2} { 2 \gamma} 
\end{array}
\end{equation}
And we can find the optimal solution of $\phi =(a_1, a_2 ), \; \theta=(b_1, b_2)$ with the following loss if we have samples $\{\tilde x_n, n=1, \ldots, N \} \sim \mathcal N \left ( 
\left [
\begin{array}{c}
0\\
0
\end{array}
\right ]
, 
\left [
\begin{array}{cc}
1 + \gamma & 2\\
2 & 4 + \gamma
\end{array}
\right ]
\right )$.

\begin{equation}
\label{eq:loss-in-detail-penalty-vae-gamma}
\begin{array}{ll}
\underset {\phi,\theta}  {\min  } \;  L(\mathbf \theta,  \mathbf \phi)  =  \frac 1 N  \sum_{n=1}^N [ \frac {( a_1 \tilde x_{n1} + a_2 \tilde x_{n2})^2 } 2 + \frac {b_1^2 + b_2^2 } {2 ( 5 + \gamma) } + \frac{ (b_1(a_1\tilde x_{n1} + a_2 \tilde x_{n2}) -\tilde x_{n1})^2 + (b_2(a_1\tilde x_{n1}  + a_2 \tilde x_{n2}) -\tilde x_{n2})^2} { 2  \gamma } ]
\end{array}
\end{equation}

We can solve the above optimization problem with first order condition. 
\begin{equation}
 \begin{array}{l}
\frac {\partial L} { \partial a_1} =\frac 1 N  \sum_{n=1}^N [ \tilde x_{n1} ( a_1 \tilde x_{n1} + a_2 \tilde x_{n2}) + \frac { b_1 \tilde x_{n1}(b_1(a_1\tilde x_{n1} + a_2 \tilde x_{n2}) -\tilde x_{n1}) + b_2 \tilde x_{n1} (b_2(a_1\tilde x_{n1}  + a_2 \tilde x_{n2}) -\tilde x_{n2}) } \gamma]
 = 0 \\
\frac {\partial L} { \partial a_2} =\frac 1 N  \sum_{n=1}^N [ \tilde x_{n2} ( a_1 \tilde x_{n1} + a_2 \tilde x_{n2}) + \frac { b_1 \tilde x_{n2}(b_1(a_1\tilde x_{n1} + a_2 \tilde x_{n2}) -\tilde x_{n1}) + b_2 \tilde x_{n2} (b_2(a_1\tilde x_{n1}  + a_2 \tilde x_{n2}) -\tilde x_{n2}) } \gamma ]
 = 0 \\
\frac {\partial L} { \partial b_1} = \frac 1 N  \sum_{n=1}^N [ \frac {b_1 } { 5 + \gamma } +  \frac {   (a_1\tilde x_{n1} + a_2 \tilde x_{n2}) (b_1(a_1\tilde x_{n1} + a_2 \tilde x_{n2}) -\tilde x_{n1}) } \gamma  ]=0\\
\frac {\partial L} { \partial b_2} =\frac 1 N  \sum_{n=1}^N [ \frac {b_2 } { 5 + \gamma } +  \frac {   (a_1\tilde x_{n1} + a_2 \tilde x_{n2}) (b_2(a_1\tilde x_{n1} + a_2 \tilde x_{n2}) -\tilde x_{n2}) } \gamma  ]= 0 
 \end{array}
\end{equation}
Now, we use the approximation $\frac 1 N \sum_{n=1}^N \tilde x_{n1}^2 \approx \mathbb E \tilde x_1^2 = 1 + \gamma $, \; $\frac 1 N \sum_{n=1}^N \tilde x_{n2}^2 \approx \mathbb E \tilde x_1^2 = 4 + \gamma$,\;$\frac 1 N \sum_{n=1}^N \tilde x_{n1}  \tilde x_{n2} \approx \mathbb E \tilde x_1\tilde x_2 =2$ into the above equation, we get 
\begin{equation}
 \begin{array}{l}
 ( 1+ \gamma ) a_1 + 2a_2 +  \frac {b^2_1[ ( 1+ \gamma ) a_1 + 2 a_2 ] - ( 1+ \gamma )b_1  + b^2_2[ ( 1+ \gamma ) a_1 + 2a_2 ] -2b_2 } \gamma
  \approx 0 \\
 2a_1  + (4+\gamma)a_2  +  \frac { b^2_1[2a_1 + (4 + \gamma ) a_2]  -2b_1 + b^2_2 [ 2a_1  + ( 4 + \gamma ) a_2 ]  - (4+\gamma) b_2 } \gamma
  \approx 0 \\
 \frac {b_1 } { 5 + \gamma } +  \frac { b_1 [ ( 1 + \gamma) a_1^2 + ( 4 + \gamma ) a_2^2 +4a_1a_2)  - [(1 + \gamma) a_1 + 2 a_2)]  } \gamma
 \approx 0\\
 \frac {b_2 } { 5 + \gamma }  +  \frac { b_2 [ ( 1 + \gamma )  a_1^2 +  ( 4 + \gamma )  a_2^2 + 4a_1a_2)  -[2a_1 + ( 4+ \gamma ) a_2]} \gamma \approx 0 
 \end{array}
\end{equation}
I can not solve the above equation with explicit solution since it will evolve a five order equation about $a_1, a_2$ when eliminating out $b_1,b_2$. But we can check that the optimal solution $\phi^\star = (a^\star_1, a^\star_2) = (\frac 1 { 5 + \gamma }, \frac 2 {5 + \gamma} )$, $\theta^\star = (b^\star_1, b^\star_2) = (1, 2)$ satisfies the above stable condition exactly. This shows that the variational auto-encoder has the power to find the true solution. Note that when $\tilde x \sim  \mathcal N \left ( 
\left [
\begin{array}{c}
0\\
0
\end{array}
\right ]
, 
\left [
\begin{array}{cc}
1+ \gamma & 2\\
2 & 4+\gamma
\end{array}
\right ]
\right )$, 
the $\mu(\tilde x,\phi^\star) = \frac 1 { 5+ \gamma}  \tilde x_1 + \frac 2 { 5+ \gamma}   \tilde x_2 \sim \mathcal N(0, \frac 5 { 5 + \gamma })$ is close to the true hidden variable distribution of $z \sim  \mathcal N(0, 1)$ if the variance  $\gamma$ of the noise trends to zero.  Note  that $z = \mu(\tilde x,\phi^\star)  + \epsilon_z,\; \epsilon_z \sim \mathcal N(0, \frac{\gamma} {5 + \gamma})$ which means that the noise level in the $x$ space influence on the hidden variable $z$ space, and the impact of noise on the hidden variable is  to be proportional with the noise level on the $x$ space. 
The $\mu(z,\theta^\star) = \left [
\begin{array}{c}
z\\
2z
\end{array}
\right ]$  capture the true distribution of the data. 
 From the relation  $\tilde x = \mu(z,\theta^\star)  + \epsilon_x = x + \epsilon_x, \;  \epsilon_x \sim  \mathcal N \left ( 
\left [
\begin{array}{c}
0\\
0
\end{array}
\right ]
, 
\left [
\begin{array}{cc}
\gamma& 0\\
0 & \gamma
\end{array}
\right ]
\right )$, we see that we can use the variational auto-encoder to denoise the data, we use the mean of the posterior distribution $p_{\theta^\star}(x|z)$ to model the distribution of the data, i.e., $p_{\theta^\star}(x) = \int_{z \;| \;x \approx \mathbb E_{ \tilde x \sim p_{\theta^\star}(\tilde x|z)  } \tilde x}  p_{\theta^\star}(z) dz$.

\section{scVI: single cell variational inference}
\begin{figure}
  \centering
  \includegraphics[width=150mm]{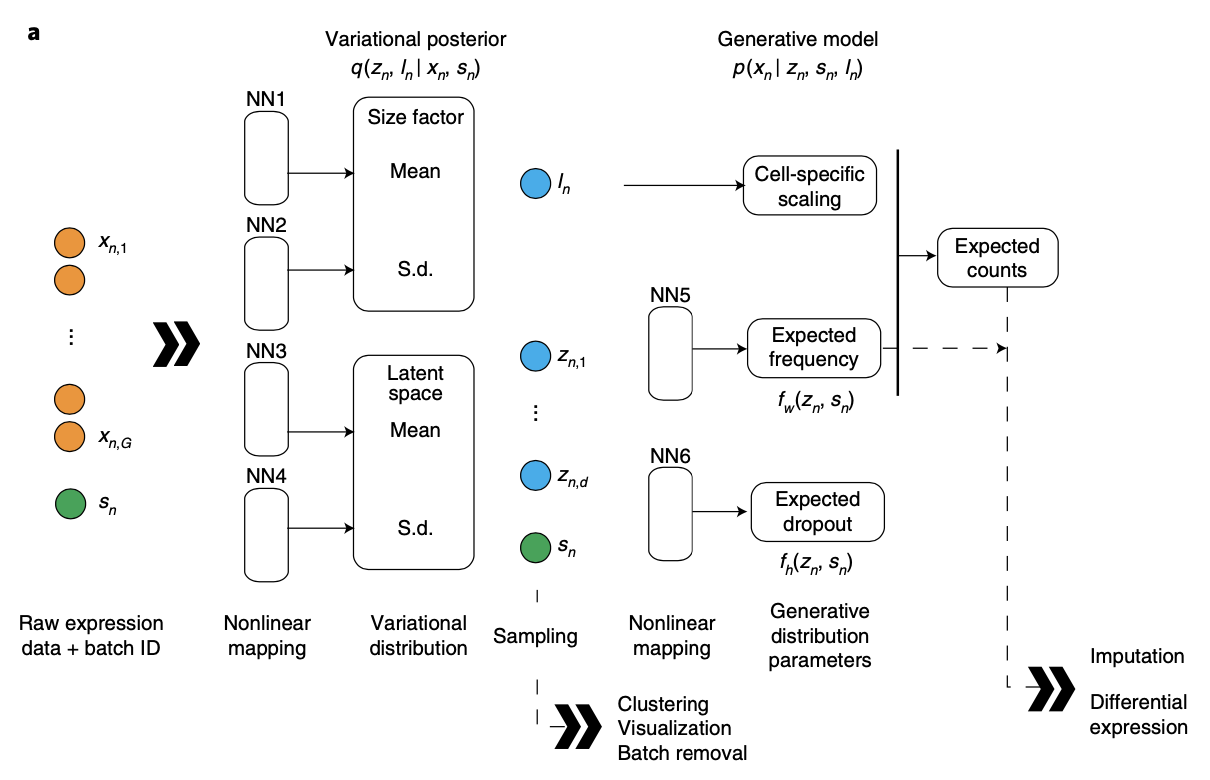}
   \caption{overview of scVI. Given a gene expression matrix with batch annotations as input, scVI learns a nonlinear embedding of the cells that can be used for multiple analysis tasks.  The neural networks used to compute the embedding and the distribution of gene expression. NN, neural network. $f_w$ and $f_h$ are functional representations of NN5 and NN6, respectively. }
\label{fig:scvi-model} 
\end{figure}

The above arguments basically introduce the work principle of the variational auto-encoder. Now we arrived the main goal, to  introduce the mathematical model of scVI(\cite{scVI}). Let we first get sense of the model with the graphical abstract in  Figure~\ref{fig:scvi-model}.

We assume that the expression $x_{ng}$ where $n$ the index of cell, and $g$ is the index gene can be generated by the following process, which characterize the probability distribution of  expression. 
\begin{equation}
\label{eq:scvi-generation}
\begin{array}{c}
z_n \sim \text{ Normal}(0, I)\\
l_n  \sim \text{ log normal}( l_{\mu} , l_{\sigma^2})\\
\rho_n = f_w(z_n, s_n) \\ 
w_{ng} \sim \text{Gamma}( \theta^g,  \frac {\theta^g} {\rho_n^g} ) \\ 
y_{ng} \sim \text{Poisson}(l_n w_{ng})\\
h_{ng} \sim \text{Bernoulli}( \frac 1 { 1 + \exp( - f^g_h(z_n, s_n))} )\\
x_{ng} = \left \{
	\begin{array} {c}
	y_{ng} \; \text{if }  h_{ng} = 0\\ 
	0 \; \text{otherwise}  
	\end{array}
	\right .
\end{array}
\end{equation}
From the above generation process,
$z_n \sim \mathcal N(\mathbf 0, \mathbf I)$ is a standard $d$-dimensional  normal distribution, which has a pdf $\frac 1 {(2\pi)^{\frac d 2}} \exp( -\frac 1 2 ||z_n||_2^2)$.  $l_n \sim \log \text{Normal}(l_\mu, l_{\sigma^2})$ is a log normal distribution with pdf $ \frac{1}{\sqrt{2\pi l_{\sigma^2}} l_n } \exp( - \frac { ( \log l_n - l_u)^2}{2 l_{\sigma^2}})$ which has mean $ \exp(l_\mu + l_{\sigma^2}/2)$ and variance $(e^{l_{\sigma^2}} -1 ) e^{2 l_\mu + l_{\sigma^2}}$. $w_{ng} \sim \text{Gamma}( \theta^g,  \frac {\theta^g} {\rho_n^g} ) $ is a Gamma distribution with shape parameter $\theta^g$ and rate parameter $ \frac {\theta^g} {\rho_n^g} $, its pdf is given by $\frac{ \left ( \frac {\theta^g} {\rho_n^g} \right )^{\theta^g}   }{\Gamma(\theta^g)} w_{ng}^{\theta^g - 1} \exp(- \frac {\theta^g} {\rho_n^g}w_{ng}) I_{w_{ng} \ge 0}$ which has mean $\rho_n^g$ and variance $(\rho_n^g)^2/\theta^g$, where the gamma function is defined by $\Gamma(\alpha) = \int_0^{\infty} x^{\alpha - 1} e^{- x} dx$. The Poisson variable $y_{ng} \sim \text{Poisson}(l_n w_{ng})$ has discrete distribution function $P(y_{ng} = k ) = \frac {(l_n w_{ng})^k}{k!} \exp(-l_n w_{ng}), \; k= 0,1, \ldots$ with an equal variance  and mean $l_n w_{ng}$. The Bernoulli random variable $h_{ng} \sim \text{Bernoulli}(\frac 1 { 1 + \exp{ - f^g_h(z_n, s_n)}})$ is $\{ 0, 1\}$ valued discrete random variable with discrete distribution function $P(h_{ng} =0) = \frac { \exp( - f^g_h(z_n, s_n)  } { 1 + \exp( - f^g_h(z_n, s_n))}$ and $P(h_{ng} =1) = \frac 1 { 1 + \exp( - f^g_h(z_n, s_n))} $ which has mean $ \frac 1 { 1 + \exp( - f^g_h(z_n, s_n))})$ and variance $ \frac { \exp( - f^g_h(z_n, s_n)  } { (1 + \exp( - f^g_h(z_n, s_n)))^2}$. 

We can get  more concise distribution if we integrate out the intermediate variable. The $y_{ng}$ follows a negative binomial distribution when given the parameter $\rho_n^g, \theta, \; l_n$, and the $x_{ng}$ follows a zero inflated binomial distribution when given the parameter $\rho_n^g, \theta, \; l_n, \; f_n^g(z_n,s_n)$. 
We now give the deduction of the negative  binomial  distribution of $y_{ng}$ when we are given  $\rho_n^g, \theta, \; l_n$,  i.e, we show that Gamma-Poisson mixture will lead a negative  binomial  distribution. 
\begin{equation}
\begin{array}{ll}
P(w_{ng}|\rho_n^g, \theta^g)	& =\frac{ \left ( \frac {\theta^g} {\rho_n^g} \right )^{\theta^g}   }{\Gamma(\theta^g)} w_{ng}^{\theta^g - 1} \exp(- \frac {\theta^g} {\rho_n^g}w_{ng}) I_{w_{ng} \ge 0} dw_{ng} \\
P(y_{ng} =k  | l_n, w_{ng})		&= \frac {(l_n w_{ng})^k}{k!} \exp(-l_n w_{ng})\\
P(y_{ng} =k  | l_n,\rho_n^g, \theta^g) 	&= \int_{w_{ng}} P(y_{ng} =k  | l_n, w_{ng})  P(w_{ng}|\rho_n^g, \theta^g) \\ 
	&= \int_{w_{ng}} \frac {(l_n w_{ng})^k}{k!} \exp(-l_n w_{ng})  \frac{ \left ( \frac {\theta^g} {\rho_n^g} \right )^{\theta^g}   }{\Gamma(\theta^g)} w_{ng}^{\theta^g - 1} \exp(- \frac {\theta^g} {\rho_n^g}w_{ng}) I_{w_{ng} \ge 0} dw_{ng}  \\
	&= \frac 1 {k!\Gamma(\theta^g)} \frac{\left ( \frac {\theta^g} {\rho_n^g} \right )^{\theta^g}    l_n^k} {(\frac {\theta^g} {\rho_n^g} +l_n)^{k+\theta^g}}  \\
	&\quad  \int_0^\infty [(\frac {\theta^g} {\rho_n^g} +l_n)w_{ng}]^{k +\theta^g-1} \exp( - ( \frac {\theta^g} {\rho_n^g} +l_n)w_{ng}) d((\frac {\theta^g} {\rho_n^g} +l_n)w_{ng}) \\
	&= \frac {\Gamma(k+\theta^g)} {k! \Gamma(\theta^g)} \frac{ \left ( \frac {\theta^g} {\rho_n^g} \right )^{\theta^g}  l_n^k} {(\frac {\theta^g} {\rho_n^g} +l_n)^{k+\theta^g}}\\
	& \sim \text{NB}(size=\theta^g,\;p= \frac{l_n}{\frac {\theta^g} {\rho_n^g}  + l_n})\\
	&=  \frac {\Gamma(k+\theta^g)} { \Gamma (k+1) \Gamma(\theta^g)}   \left ( \frac {\theta^g} {\theta^g + l_n \rho_n^g} \right )^{\theta^g}   \left ( \frac {l_n \rho_n^g} {\theta^g + l_n \rho_n^g} \right )^k \\
	& \sim \text{NB}(dispersion=\theta^g,\; \mu=  l_n\rho_n^g)\\
	
\end{array}
\end{equation}
where $\text{NB}(n,p)$ is the negative binomial distribution with $size = n$ and $prob = p$ has density 
\begin{equation}
\label{eq:nb-prob}
\begin{array}{ll}
\frac{ \Gamma(x+n) }{  \Gamma(n) x!} (1 - p)^n p^x
\end{array}
\end{equation}
where $p \in (0,1]$ is the probability of success in the Bernoulli distribution, and $x = 0, 1,2 \ldots, n>0$ represents the number of success which occur in a sequence of Bernoulli trials (with the probability of success $p$)  before a target number $n$ of failure  is reached, with the probability given in equation (\ref{eq:nb-prob}).  The mean is $\mu = np/(1-p)$ and variance $ np/(1-p)^2$. An alternative parametrization (often used in ecology) is by the mean $\mu$, and size, the dispersion parameter $\theta$, where $p= \mu/(\theta+\mu)$. The variance is $\mu + \mu^2/\theta$ in this parametrization. 

This give us that $y_{ng}$ is negative binomial distribution with $size=dispersion = \theta^g,p= \frac{l_n}{\frac {\theta^g} {\rho_n^g}  + l_n}, \;\mu=  l_n\rho_n^g$ and variance $  l_n\rho_n^g +  \frac {(l_n\rho_n^g)^2} {\theta^g}$.

Now, it obvious that $x_{ng}$ obeys the zero-inflated negative binomial distribution with probability mass function given by 
\begin{equation}
\label{eq:zinb-prob}
\begin{array}{ll}
P(x_{ng} = 0|l_n, \rho_n^g, \theta^g, f_h^g) &= P(x_{ng} = 0| h_{ng} = 0, l_n, \rho_n^g, \theta^g, f_h^g)P(h_{ng} = 0 | l_n, \rho_n^g, \theta^g, f_h^g) \\
&\quad +  P(x_{ng} = 0| h_{ng} = 1, l_n, \rho_n^g, \theta^g, f_h^g)P(h_{ng} = 1 | l_n, \rho_n^g, \theta^g, f_h^g)\\
&= P(y_{ng} = 0 | l_n, \rho_n^g, \theta^g)P(h_{ng} = 0 | l_n, \rho_n^g, \theta^g, f_h^g) \\
&\quad +  P(h_{ng} = 1 | l_n, \rho_n^g, \theta^g, f_h^g)\\
&=   \left ( \frac {\theta^g} {\theta^g + l_n \rho_n^g} \right )^{\theta^g}  \frac { \exp( - f^g_h(z_n, s_n)  } { 1 + \exp( - f^g_h(z_n, s_n))} + \frac { 1 } { 1 + \exp( - f^g_h(z_n, s_n))} \\
&= \exp(S(- f_h^g + \theta^g \log \frac { \theta^g} { \theta^g + l_n\rho_n^g} - S(- f_h^g))\\
P(x_{ng} = k |l_n, \rho_n^g, \theta^g, f_h^g, k>0) &=  P(x_{ng} = k| h_{ng} = 0, l_n, \rho_n^g, \theta^g, f_h^g,k>0)P(h_{ng} = 0 | l_n, \rho_n^g, \theta^g, f_h^g)\\
	&=  P(y_{ng} = k| l_n, \rho_n^g, \theta^g)P(h_{ng} = 0 |  f_h^g)\\
	&=  \frac {\Gamma(k+\theta^g)} { \Gamma (k+1) \Gamma(\theta^g)}   \left ( \frac {\theta^g} {\theta^g + l_n \rho_n^g} \right )^{\theta^g}   \left ( \frac {l_n \rho_n^g} {\theta^g + l_n \rho_n^g} \right )^k  \frac { \exp( - f^g_h(z_n, s_n)  } { 1 + \exp( - f^g_h(z_n, s_n))} \\
	&= \exp( -f_h^g - S(-f_h^g) + \theta^g \log \frac { \theta^g} { \theta^g + l_n\rho_n^g} \\
	&\quad   + k  \log  \frac k {\theta^g + l_n\rho_n^g}  + \log  \frac {\Gamma(k+\theta^g)} { \Gamma (k+1) \Gamma(\theta^g)}  )
\end{array}
\end{equation}
where $S(x) := \log(1 + e^x)$ is the softplus function. The above probability mass function can be represented in a compact form. 
\begin{equation}
\label{eq:zinb-prob-compact}
\begin{array}{ll}
P(x_{ng} | \phi,z_n,s_n)&=\mathbb 1_{x_{ng} =0} \exp ( S(- f_h^g +\theta^g \log \frac { \theta^g} { \theta^g + l_n\rho_n^g}  - S(- f_h^g)  )\\
&+ \mathbb 1_{x_{ng} >0}  \exp( -f_h^g - S(-f_h^g) + \theta^g \log \frac { \theta^g} { \theta^g + l_n\rho_n^g}\\
	&\quad    + x_{ng}  \log  \frac { x_{ng} } {\theta^g + l_n\rho_n^g}  + \log  \frac {\Gamma(x_{ng}+\theta^g)} { \Gamma (x_{ng}+1) \Gamma(\theta^g)}  )
\end{array}
\end{equation}

\begin{figure}
  \centering
  \includegraphics[width=139mm]{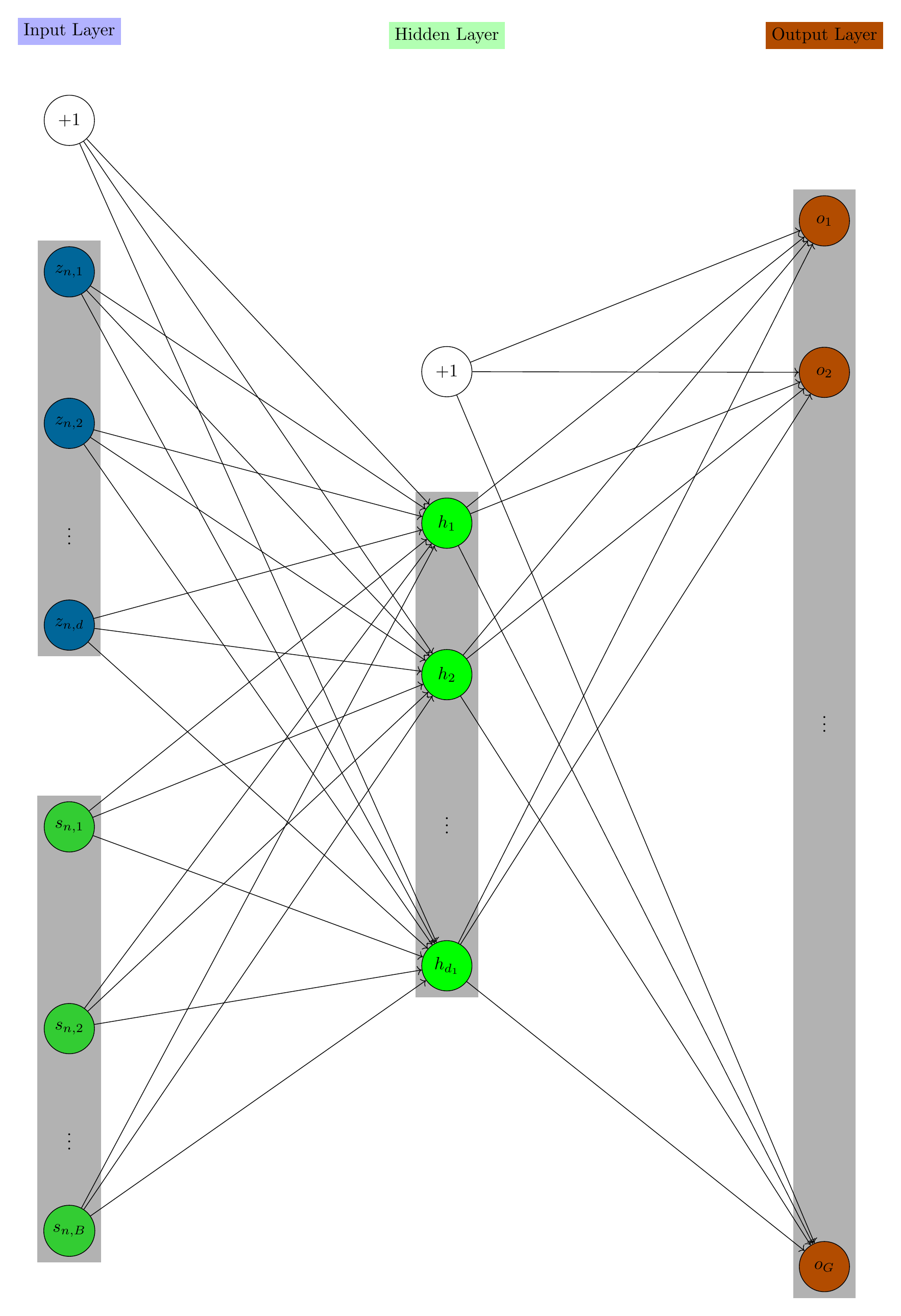}
   \caption{The one-layer MLP of representation of $f_w(z_n,s_n)$ with hidden neuron $ h_i = \text{ReLU}(\sum_{j= 1}^dW^{(z)}_{i,j} z_{n,j} + \sum_{j = 1}^BW^{(s)}_{i,j} s_{n,j} +b_i), \; i= 1, \ldots, d_1$  and output neuron  $o_g= \sum_{j= 1}^{d_1} W^{(h)}_{g,j} h_j + b^{(h)}_g,  \; g = 1, \ldots, G$, and the final expected frequency $\rho_n^g =  = \frac { o_g } {\sum_j e^{ o_j}}, \; g = 1, \ldots, G$, $f_w(z_n,s_n) = \rho_n$.}
\label{fig:nn} 
\end{figure}

To get a sense of neural networks function $f_w(z_n,s_n)$ where $z_n \in \mathbb R^d$ and $s_n \in \mathbb R^{B}$ is one-hot representation of the batch id of the cell $n$, we plot the following cartoon figure~\ref{fig:nn} to represent it. This network contains one hidden layer with $d_1$ neurons, $ h_i = \text{ReLU}(\sum_{j= 1}^dW^{(z)}_{i,j} z_{n,j} + \sum_{j = 1}^BW^{(s)}_{i,j} s_{n,j} +b_i), \; i= 1, \ldots, d_1$ where ReLU function is a elementwise function defined by  $ \text{ReLU}(x) = \max(x,0)$, the $W^{(z)} \in \mathbb R^{d_1 \times d}, \; W^{(s)} \in \mathbb R^{d_1 \times B}, \; b \in \mathbb R^{d_1}$ are the weights of the hidden layer. The output layer is built on the hidden layer with a linear mapping then with a Softmax function, i.e. $o_g= \sum_{j= 1}^{d_1} W^{(h)}_{g,j} h_j + b^{(h)}_g,  \; g = 1, \ldots, G$, $\rho_n^g =  = \frac { o_g } {\sum_j e^{ o_j}}, \; g = 1, \ldots, G$, where $W^{(h)} \in \mathbb R^{G \times d}, \; b^{(h)} \in \mathbb R^{G}$ are the weights of the output layer. Finally, $f_w(z_n,s_n) = \rho_n $ and $w = \{W^{(z)}, \;W^{(s)}  ,\;  b  ,\;  W^{(h)} ,\;   b^{(h)}  \}$. This simple neural networks capture the basic structure of neural networks, i.e., linear transform followed by nonlinear mapping, and cascade this basic building blocks. In the practice, the neural network can add more hidden layers to model complex dependence between the input variables and output variables. And for an efficiency and the algorithm stability, we can add batch normalization (\cite{2015batchnorm}) after the linear transform before the nonlinear mapping to make the gradients go through a numerical stable path. And we can also add the  Dropout layer  (\cite{dropout-srivastava14a})  which random drop out the connection between neurons with  some fix probability to avoid overfitting. And we can also add a l2 norm of weights on the variational lower bound to stable the algorithm, which is favorable the weights near the origin. 

As we have prepared the basic information  to understand the scVI, we now begin to view the whole picture of the scVI. As we see from figure~\ref{fig:scvi-model}. It is a  conditional variational auto-encoder, which add a conditional variable batch ID to the generation model. It is follows the same logic as the variational autoencoder with a minimal change, which is that we only need to make all the probability conditioned on the batch id variable $s$. So following the almost the same logic in the deduction of loss function of the variational auto-encoder, we now begin to deduce the variational lower bound of the scVI model. Firstly, the  log  likelihood  conditioned on the batch ID is given by $\log p_\psi(x_n|s_n)$ for cell $n$ (here and after, we use $\psi$ as the model parameter for the generation probability decoder model instead of $\theta$ above, since the $\theta$ parameter was used as the gene dispersion parameter), then we have the variational lower bound of the log-likelihood conditioned on batch ID:
\begin{equation}
\label{eq:varitional-lower-bound-scVI}
 \begin{array}{l}
 L( \psi,  \phi, x_n, s_n)= -D_{KL}(q_{\phi}(z_n, l_n | \; x_n, s_n) ||  p_{\psi}(z_n, l_n | s_n)  ) +  \mathbb E_{q_{\phi}(z_n, l_n | \;x_n, s_n)}  \log p_{\psi }(x_n | z_n, s_n, l_n)  \\
 \end{array}
\end{equation}

The $q_{\phi}(z_n, l_n | \; x_n, s_n)$ is a probability encoder which encode gene expression $x_n$ into  low dimensional hidden variable $z_n$ and the  surrogate of the library size  $l_n$ conditioned on $s_n$. The parameters $\phi$ are the collections of weight of NN1, NN2, NN3, NN4 in figure~\ref{fig:scvi-model}. The encoder consists of two sub-encoder $q_{\phi}(z_n, l_n | \; x_n, s_n) = q_{\phi}(z_n | \; x_n, s_n ) q_{\phi}( l_n | \; x_n, s_n)$. The variational distribution $q_{\phi}(z_n | \; x_n, s_n )$ is chosen to be Gaussian with a diagonal covariance, with mean given by an encoder network NN1 applied to $x_n, s_n$ and diagonal deviation( the square root of the diagonal variance ) given by the encoder network NN2 applied to $x_n, s_n$. The variation distribution $q_{\phi}( l_n | \; x_n, s_n)$ is chosen to be log-normal with scalar mean and variance, with mean and standard deviation (the square root of the variance)  given by neural network NN4, NN5, respectively. 
We apply the reparameterization trick on the variational distribution. 
\begin{equation}
\label{eq:reparameterization-scVI}
 \begin{array}{l}
z_n \sim  q_{\phi}(z_n | \; x_n, s_n ) \sim \mathcal N(   f_{NN1, \phi} ( x_n, s_n ) , \text{diag}\{( f_{NN2, \phi}( x_n, s_n ))^2 \}) \\
z_n =  f_{NN1, \phi}( x_n, s_n ) +  f_{NN2, \phi}( x_n, s_n ) \epsilon_z,  \epsilon_z \sim \mathcal N(0 ,I_d) \\
l_n \sim   q_{\phi}(l_n | \; x_n, s_n ) \sim  \text{log normal }(  f_{NN3, \phi}( x_n, s_n ), ( f_{NN4, \phi}( x_n, s_n ))^2 ) \\
l_n = \exp(  f_{NN3, \phi}( x_n, s_n ) +  f_{NN4, \phi}( x_n, s_n ) \epsilon_l), \;  \epsilon_l \sim \mathcal N(0,1)
 \end{array}
\end{equation}
The prior $p_{\psi}(z_n, l_n | s_n) = p(z_n | s_n) p(l_n | s_n)$ is chosen as a fixed distribution, where $p(z_n | s_n) \sim \mathcal N(0, I_d)$ and $p(l_n | s_n) \sim \text{log normal}(\mu_{b_n}, \sigma^2_{b_n} )$ where $b_n$ is the batch ID of cell $n$ and 
\begin{equation}
 \begin{array}{ll}
\mu_{b_n} := \frac{\sum_{i  \in \text {batch } b_n } \log (\sum_{g=1}^G x_{i,g}) } {\sum_{i  \in \text {batch } b_n} 1} \\
\sigma^2_{b_n}  :=  \frac{(\sum_{i  \in \text {batch } b_n} \log (\sum_{g=1}^G x_{i,g})  - \mu_{b_n})^2} {\sum_{i  \in \text {batch }b_n } 1}
 \end{array}
\end{equation}, 
i.e. the sample log mean and variance of the log of  library size of cells in batch $b_n$. 

To get the analytical expression of the KL divergence, we first calculate a simple example of $D_{KL}(\mathcal N(m_1, \sigma_1^2) || \mathcal N(m_1, \sigma_1^2))$. 
\begin{equation}
\label{eq:KL-example-normal-distribution}
 \begin{array}{ll}
 D_{KL}(\mathcal N(m_1, \sigma_1^2) || \mathcal N(m_2, \sigma_2^2)) 	& = \int_{-\infty}^{\infty} \frac 1 {2\pi \sigma_1^2} \exp(- \frac {(x- m_1)^2}{ 2 \sigma_1^2} ) \log \frac { \frac 1 {2\pi \sigma_1^2} \exp(- \frac {(x- m_1)^2}{ 2 \sigma_1^2}) } { \frac 1 {2\pi \sigma_2^2} \exp(- \frac {(x- m_2)^2}{ 2 \sigma_2^2} )} dx \\
 	&= \int_{-\infty}^{\infty} \frac 1 {2\pi \sigma_1^2} \exp(- \frac {(x- m_1)^2}{ 2 \sigma_1^2})  [\log \frac {\sigma_2}{\sigma_1} - \frac {(x- m_1)^2}{ 2 \sigma_1^2}   + \frac {(x- m_2)^2}{ 2 \sigma_2^2} ]  dx \\
	 &= \log \frac {\sigma_2}{\sigma_1}  - \frac 1 2 + \frac { \sigma_1^2 + (m_1 - m_2)^2} { 2 \sigma^2_2}
 \end{array}
\end{equation}
So we have 
\begin{equation}
\label{eq:KL-zn}
 \begin{array}{ll}
D_{KL}(q_{\phi}(z_n | \; x_n, s_n )  ||  p(z_n)) 	&= D_{KL} ( \mathcal N(   f_{NN1, \phi} ( x_n, s_n ) , \text{diag}\{( f_{NN2, \phi}( x_n, s_n ))^2 \})  ||  \mathcal N(0, I_d)) \\
	&= \sum_{i=1}^d  [\log \frac 1 {[ f_{NN2, \phi}( x_n, s_n )]_i }  - \frac 1 2 + \frac {[ f_{NN2, \phi}( x_n, s_n )]_i^2 +[   f_{NN1, \phi} ( x_n, s_n )]_i^2 }  {2} ] 
 \end{array}
\end{equation}
If $Y_1 \sim \text{log normal}(\mu_1, \sigma_1^2)$ with pdf $p_1(y)$  and $Y_2 \sim \text{log normal}(\mu_2, \sigma_2^2)$ with pdf $p_2(y)$ , then $\log Y_1 \sim \mathcal N(m_1, \sigma_1^2)$  with pdf $p_1(e^x)e^x$ and $\log Y_2 \sim  \mathcal N(m_1, \sigma_1^2)$ with pdf $p_1(e^x)e^x$. 
\begin{equation}
\label{eq:KL-example-log-normal}
 \begin{array}{ll}
D_{KL}(   \text{log normal}(\mu_1, \sigma_1^2) || \text{log normal}(\mu_2, \sigma_2^2) ) 	&= \int_{y=0}^{\infty} p_1(y) \log \frac{p_1(y)} {p_2(y)} dy\\
	&= \int_{x=-\infty}^{\infty}  p_1(e^x) e^x\log \frac{p_1(e^x)e^x} {p_2(e^x)e^x} dx\\
	&= D_{KL}(\mathcal N(m_1, \sigma_1^2) || \mathcal N(m_2, \sigma_2^2))  \\
	& =  \log \frac {\sigma_2}{\sigma_1}  - \frac 1 2 + \frac { \sigma_1^2 + (m_1 - m_2)^2} { 2 \sigma^2_2}
\end{array}
\end{equation}
So we have 
\begin{equation}
\label{eq:KL-ln}
 \begin{array}{ll}
D_{KL}(  q_{\phi}(l_n | \; x_n, s_n )  ||  p(l_n)) ) 	&=D_{KL}(   \text{log normal }(  f_{NN3, \phi}( x_n, s_n ), ( f_{NN4, \phi}( x_n, s_n ))^2 )   || \text{log normal}( \mu_{b_n}, \sigma_{b_n}^2) )  \\
	&=   \log \frac {\sigma_{b_n}}{ f_{NN4, \phi}( x_n, s_n )}  - \frac 1 2 + \frac { ( f_{NN4, \phi}( x_n, s_n ))^2 + ( f_{NN3, \phi}( x_n, s_n ) - m_{b_n})^2} { 2 \sigma_{b_n}^2}
\end{array}
\end{equation}
Combining equation (\ref{eq:KL-zn}) and (\ref{eq:KL-ln}), we get
\begin{equation}
\label{eq:KL-total}
 \begin{array}{ll}
D_{KL}(q_{\phi}(z_n, l_n | \; x_n, s_n) ||  p_{\psi}(z_n, l_n)  ) 	&=D_{KL}(  q_{\phi}(z_n | \; x_n, s_n ) q_{\phi}(l_n | \; x_n, s_n )  || p(z_n)  p(l_n)) )  \\
	&=  D_{KL}(q_{\phi}(z_n | \; x_n, s_n )  ||  p(z_n)) D_{KL}(  q_{\phi}(l_n | \; x_n, s_n )  ||  p(l_n)) ) \\
	&= \sum_{i=1}^d  [\log \frac 1 {[ f_{NN2, \phi}( x_n, s_n )]_i }  - \frac 1 2 + \frac {[ f_{NN2, \phi}( x_n, s_n )]_i^2 +[   f_{NN1, \phi} ( x_n, s_n )]_i^2 }  {2} ] \\
	&\quad + \log \frac {\sigma_{b_n}}{ f_{NN4, \phi}( x_n, s_n )}  - \frac 1 2 + \frac { ( f_{NN4, \phi}( x_n, s_n ))^2 + ( f_{NN3, \phi}( x_n, s_n ) - m_{b_n})^2} { 2 \sigma_{b_n}^2} 
\end{array}
\end{equation}

We now only need the reconstruction error term to get the final computable objective function. We use the reparameterization trick on the reconstruction error term and use sample average to estimate the expectation. 
\begin{equation}
\label{eq:reconstruction-error}
 \begin{array}{ll}
 \mathbb E_{q_{\phi}(z_n, l_n | \;x_n, s_n)}  \log p_{\psi }(x_n | z_n, s_n, l_n) 	\\
 = \mathbb E_{\epsilon_z \sim \mathcal N(0, I_d) ,\; \epsilon_l \sim \mathcal N(0,1) }  \log p_{\psi }(x_n |   f_{NN1, \phi}( x_n, s_n ) +  f_{NN2, \phi}( x_n, s_n ) \epsilon_z, s_n, \\
  \quad \exp(  f_{NN3, \phi}( x_n, s_n ) +  f_{NN4, \phi}( x_n, s_n ) \epsilon_l))  \\
 \approx  \frac 1 K \sum_{k=1}^K \log p_{\psi }(x_n |   f_{NN1, \phi}( x_n, s_n ) +  f_{NN2, \phi}( x_n, s_n ) \epsilon_z^{(k)}, s_n,\\
 \quad  \exp(  f_{NN3, \phi}( x_n, s_n ) +  f_{NN4, \phi}( x_n, s_n ) \epsilon_l^{(k)}))   \\
 \text{where } \epsilon_z^{(k)} \overset{i.i.d}{ \sim} \mathcal N(0, I_d), \; \epsilon_l^{(k)} \overset{i.i.d}{ \sim} \mathcal N(0, 1)
\end{array}
\end{equation}

Substituting equation (\ref{eq:KL-total}) and  (\ref{eq:reconstruction-error})  into equation (\ref{eq:varitional-lower-bound-scVI}), we get the computable variational lower bound of the log-likelihood of cell $n$.
\begin{equation}
\label{eq:variational-lower-bound-cell-n}
 \begin{array}{ll}
  L( \psi,  \phi, x_n, s_n)&\approx  \tilde L( \psi,  \phi, x_n, s_n) \\
	&:= \sum_{i=1}^d  [\log \frac 1 {[ f_{NN2, \phi}( x_n, s_n )]_i }  - \frac 1 2 + \frac {[ f_{NN2, \phi}( x_n, s_n )]_i^2 +[   f_{NN1, \phi} ( x_n, s_n )]_i^2 }  {2} ] \\
	&\quad	+ \log \frac {\sigma_{b_n}}{ f_{NN4, \phi}( x_n, s_n )}  - \frac 1 2 + \frac { ( f_{NN4, \phi}( x_n, s_n ))^2 + ( f_{NN3, \phi}( x_n, s_n ) - m_{b_n})^2} { 2 \sigma_{b_n}^2} \\
	&\quad + \frac 1 K \sum_{k=1}^K \log p_{\psi }(x_n |  z_n^{(k)} , s_n,  l_n^{(k)} )   \\
	&= \sum_{i=1}^d  [\log \frac 1 {[ f_{NN2, \phi}( x_n, s_n )]_i }  - \frac 1 2 + \frac {[ f_{NN2, \phi}( x_n, s_n )]_i^2 +[   f_{NN1, \phi} ( x_n, s_n )]_i^2 }  {2} ] \\
	&\quad 	+ \log \frac {\sigma_{b_n}}{ f_{NN4, \phi}( x_n, s_n )}  - \frac 1 2 + \frac { ( f_{NN4, \phi}( x_n, s_n ))^2 + ( f_{NN3, \phi}( x_n, s_n ) - m_{b_n})^2} { 2 \sigma_{b_n}^2} \\
	&\quad  + \frac 1 K \sum_{k=1}^K  \sum_{g=1}^G \{
 \mathbb 1_{x_{ng} =0} [S(- f_h^g(z_n^{(k)}, s_n) +\theta^g \log \frac { \theta^g} { \theta^g + l_n^{(k)} f_w^g(z_n^{(k)}, s_n)}  - S(- f_h^g(z_n^{(k)}, s_n))] \\
	&\quad  \mathbb 1_{x_{ng} >0}  [  -f_h^g(z_n^{(k)}, s_n) - S(-f_h^g(z_n^{(k)}, s_n)) + \theta^g \log \frac { \theta^g} { \theta^g + l_n^{(k)}f_w^g(z_n^{(k)}, s_n)}\\
	&\quad + x_{ng}  \log  \frac { x_{ng} } {\theta^g + l_n^{(k)} f_w^g(z_n^{(k)}, s_n)}  + \log  \frac {\Gamma(x_{ng}+\theta^g)} { \Gamma (x_{ng}+1) \Gamma(\theta^g)} ]\} \\
 &\text{where }  z_n^{(k)}= f_{NN1, \phi}( x_n, s_n ) +  f_{NN2, \phi}( x_n, s_n ) \epsilon_z^{(k)}, \\
 &l_n^{(k)} = \exp(  f_{NN3, \phi}( x_n, s_n ) +  f_{NN4, \phi}( x_n, s_n ) \epsilon_l^{(k)}), \\
 &\text{ and } \epsilon_z^{(k)} \overset{i.i.d}{ \sim} \mathcal N(0, I_d), \; \epsilon_l^{(k)} \overset{i.i.d}{ \sim} \mathcal N(0, 1)
\end{array}
\end{equation}
where $\psi$ is the collection of the parameters of the neural networks NN5, NN6 and $\theta$, and $\phi$ is  the collection of the parameters of the neural networks NN1, NN2, NN3, NN4. Note that the gene dispersion parameter $\theta \in \mathbb R_{+}^G$ is constant for each gene $g$ in figure~\ref{fig:scvi-model}, there are some variants of the choices of $\theta$, 
\begin{enumerate}
\item Gene dispersions are constant in each batch and each gene, $\theta \in  \mathbb R_{+}^{G\times B}$, where $B$ is the number of batches.
\item If the gene expression data has been annotated. Gene dispersions can be chosen to a constant in each class and each gene, $\theta \in  \mathbb R_{+}^{G\times C}$, where $C$ is the number of classes. 
\item Gene dispersions are chose to be specific for each cell and each gene, $\theta  \in  \mathbb R_{+}^{G\times N}$, where $N$ is number of cells in the data. In this case, $\theta$ can be models as a neural network $\theta = f_{w_\theta}(x_n,s_n)$ where $w_\theta$ is the parameters of the neural network.
\end{enumerate}

Now we can use the stochastic optimization method to maximize the objective function 
\begin{equation}
\label{eq:total-loss}
\begin{array}{l}
\underset{\psi,\phi}{ \max} \frac 1 N \sum_n   \tilde L( \psi,  \phi, x_n, s_n) 
\end{array}
\end{equation}
where $\tilde L( \psi,  \phi, x_n, s_n) $ is defined in equation (\ref{eq:variational-lower-bound-cell-n}). 

\section{ Conclusion}
Now, I have finished to introduce you the mathematical model of the scVI(\cite{scVI}), and  you can  get more details about the numerical experiments in scVI(\cite{scVI}). Since the sequencing data has accumulated a huge amount, the neural networks based model(e.g. variational auto-encoder) gives us a power to hand such a huge amount of data. However, the neural network is a black box, which means  we can hard to gains insights from its  millions parameters. This field need a large amount of exploration. Thank you for you reading, have a nice day!

\bibliographystyle{apalike}  
\bibliography{./bib/references}  

\end{document}